\documentclass[a4paper,twocolumn,11pt]{quantumarticle}
\pdfoutput=1
\usepackage[utf8]{inputenc}
\usepackage[english]{babel}
\usepackage[T1]{fontenc}
\usepackage{amsmath}
\usepackage{braket}
\usepackage{float}
\usepackage{hyperref}
\usepackage{caption}
\usepackage{tikz}
\usepackage{multirow}
\usepackage{lipsum}
\usepackage{pythonhighlight}
\usepackage{amsfonts}
\usepackage{booktabs}
\usepackage{listings}
\usepackage{subcaption}
\usepackage{tikzit}
\usepackage{pgfplots}
\usetikzlibrary{decorations.markings}
\usepackage{tikz-cd}
\usetikzlibrary{patterns.meta}
\tikzstyle{hbox}=[draw=black, shape=rectangle, fill=yellow, minimum size=.55em, inner sep=0.15em, scale=0.85]

\tikzstyle{Z dot}=[
  draw=black,
  shape=circle,
  fill={rgb,255:red,181;green,215;blue,181},
  inner sep=0.7mm,
  minimum size=0pt
]

\tikzstyle{black node}=[draw=black, shape=circle, scale=0.3, fill=black, font={\footnotesize}]

\tikzstyle{X dot}=[
  draw=black,
  shape=circle,
  fill={rgb,255:red,215;green,96;blue,96},
  inner sep=0.7mm,
  minimum size=0pt
]
\tikzstyle{midArrow}=[postaction=decorate, decoration={markings, mark=at position 0.5 with {\arrow{stealth[length=6.4pt, sep=-2pt]}}}]
\tikzstyle{sLabel}=[font={\scriptsize}, auto]
\tikzstyle{NLO}=[-, line width=.6pt, tikzit draw={rgb,255: red,117; green,44; blue,83}]
\tikzstyle{N}=[-, line width=0.7pt]
\tikzstyle{NStream}=[-, N, midArrow]
\tikzstyle{LOWire}=[-, very thin, tikzit draw={rgb,255: red,134; green,42; blue,43}]
\tikzstyle{dashedE}=[-, dashed]
\tikzstyle{LOStream}=[-, NLO, midArrow, tikzit draw={rgb,255: red,139; green,43; blue,126}]

\definecolor{codegreen}{rgb}{0,0.6,0}
\definecolor{codegray}{rgb}{0.5,0.5,0.5}
\definecolor{codepurple}{rgb}{0.58,0,0.82}
\definecolor{backcolour}{rgb}{0.95,0.95,0.92}

\lstdefinestyle{mystyle}{
	backgroundcolor=\color{backcolour},
	commentstyle=\color{codegreen},
	keywordstyle=\color{magenta},
	numberstyle=\tiny\color{codegray},
	stringstyle=\color{codepurple},
	basicstyle=\ttfamily\footnotesize,
	breakatwhitespace=false,
	breaklines=true,
	captionpos=b,
	keepspaces=true,
	numbers=left,
	numbersep=5pt,
	showspaces=false,
	showstringspaces=false,
	showtabs=false,
	tabsize=2
}
\begin{document}

\title{Optyx: A ZX-based Python library for networked quantum architectures}

\author{Mateusz Kupper}
\affiliation{Quantinuum, 17 Beaumont Street, Oxford, OX1 2NA, United Kingdom}
\affiliation{University of Sussex, United Kingdom}
\email{mateusz.kupper@quantinuum.com}
\author{Richie Yeung}
\affiliation{Quantinuum, 17 Beaumont Street, Oxford, OX1 2NA, United Kingdom}
\affiliation{University of Oxford, United Kingdom}
\author{Boldizs\'ar Po\'or}
\affiliation{Quantinuum, 17 Beaumont Street, Oxford, OX1 2NA, United Kingdom}
\affiliation{University of Oxford, United Kingdom}
\author{Alexis Toumi}
\affiliation{DisCoPy, France}
\author{William Cashman}
\affiliation{University of Oxford, United Kingdom}
\author{Giovanni de Felice}
\affiliation{Quantinuum, 17 Beaumont Street, Oxford, OX1 2NA, United Kingdom}

\maketitle

\begin{abstract}
  Distributed, large-scale quantum computing will need architectures that combine matter-based qubits with photonic links, but today's software stacks target either gate-based chips or linear-optical devices in isolation. We introduce \texttt{Optyx}, an open-source Python framework offering a unified language to program, simulate, and prototype hybrid, networked systems: users create experiments that mix qubit registers, discrete-variable photonic modes, lossy channels, heralded measurements, and real-time feedback; \texttt{Optyx} compiles them via ZX/ZW calculus into optimised tensor-network forms, and executes with state-of-the-art contraction schedulers based on \texttt{Quimb} and \texttt{Cotengra}. Benchmarking on exact multi-photon circuit simulations shows that, versus permanent-based methods, tensor network contraction can deliver speedups of orders of magnitude for low-depth circuits and entangled photon sources, and natively supports loss and distinguishability -- establishing it as both a high-performance simulator and a rapid-prototyping environment for next-generation photonic-network experiments.
\end{abstract}

\section{Introduction}
The promise of universal, fault-tolerant quantum computation is intimately tied to our ability to scale beyond the confines of a single, monolithic chip. Networked and modular architectures -- in which physically distinct quantum nodes are coherently linked by photonic interconnects -- are emerging as a compelling path to scalability \cite{main2025distributed}. By decoupling the generation, processing, and storage of quantum information, these hybrid platforms leverage the long-range, room-temperature propagation of photons to mediate entanglement between otherwise monolithic hardware, from superconducting qubits to trapped ions and neutral atoms. Recent demonstrations of remote Bell pairs \cite{main2025distributed, AghaeeRad2025, QIU2025351, PhysRevLett.130.050803}, reconfigurable photonic routers \cite{main2025distributed, AghaeeRad2025}, and metropolitan-scale quantum key distribution \cite{chen2021twin0field} highlight the rapid experimental progress. Yet translating these advances into practical, large-scale applications demands software abstractions that can express the concurrent, distributed, and multi-modal nature of networked experiments.

Current quantum programming frameworks remain largely committed to a circuit paradigm that assumes a single register of qubits evolving under a sequence of unitary gates and computational-basis measurements. Toolkits such as \texttt{Qiskit}~\cite{qiskit2024}, \texttt{Cirq}~\cite{Cirq_Developers_2025}, \texttt{tket}~\cite{Sivarajah_2021}, excel at describing gate-based processors, but offer only rudimentary support for different modes of computation such as with light-matter interaction. Conversely, photonic-native simulators focus on either Gaussian (continuous-variable) photonic computing (\texttt{Strawberry Fields}~\cite{strawberryfields}, \texttt{The Walrus}~\cite{gupt_walrus:_2019}), bosonic discrete-variable modes and permanent-based sub-routines for simulation (\texttt{Perceval}~\cite{Heurtel2023percevalsoftware}, \texttt{SOQCS}~\cite{OSCA2024101603}, \texttt{Linopt}~\cite{linopt}), or hybrid continuous-variable and discrete-variable photonics or general quantum mechanics (\texttt{Piquasso}~\cite{Kolarovszki2025piquassophotonic}, \texttt{Mr~Mustard}~\cite{mrmustard}, \texttt{QuTiP}~\cite{johansson_qutip:_2012}, \texttt{QuantumOptics.jl}~\cite{kramer_quantumoptics.jl:_2018}). The Python package \texttt{GraphiQ}~\cite{Lin2024graphiqquantum} is designed to produce quantum circuits that generate specific photonic graph states, while \texttt{Graphix}~\cite{Sunami:2022tub} provides tools for generating, optimising, and simulating measurement patterns in measurement-based quantum computing. Therefore, the existing software lacks facilities for integrating qubits, real-time feedback, or a high-level description of network topologies. In short, no existing platform offers a unified language for programming all the components of a distributed, heterogeneous architecture.

String-diagrammatic calculi such as \emph{ZX} and \emph{ZW} provide a typed, compositional representation for quantum circuits and channels, with a small set of sound rewrite rules that support semantics-preserving optimisation -- compiling via ZX enables graph-theoretic simplifications (local complementation, pivoting) and reliable circuit extraction \cite{Backens2021therebackagain}, yielding depth and gate-count reductions and, in the fault-tolerant regime, substantial $T$-count savings via phase teleportation~\cite{duncan_graph-theoretic_2020, de_beaudrap_fast_2020}. ZX-calculus can also be used to check the equivalence of quantum circuits~\cite{peham_equivalence_2022, duncan_verifying_2014}. Mature tooling further makes these calculi attractive for software development: \texttt{PyZX} (and its Rust port \texttt{QuiZX}) implements large-scale automated rewriting and validation workflows~\cite{kissinger_pyzx:_2020}. \texttt{DisCoPy}~\cite{de_felice_discopy:_2021} provides a Pythonic framework for constructing and reasoning about string diagrams, supporting the definition of custom categories and functors, including user-defined interpretations to concrete semantics such as tensor networks.

Tensor-network (TN) methods have emerged as a powerful bridge between high-level circuit descriptions and low-level, hardware-specific simulations. By exploiting entanglement structure, TNs routinely push the classical simulability frontier of qubit circuits, and have been extended to Gaussian boson sampling~\cite{Oh2024, PhysRevA.108.052604}, spin-boson dynamics~\cite{PhysRevA.94.053637}, optical circuits~\cite{PhysRevA.97.062304} and open-system modelling~\cite{PhysRevB.93.075105}. The application of TNs to \emph{multi-node photonic networks} remains in its infancy: existing methods either restrict to linear-optical interferometers or require manual construction of the network graph, and none provide an end-to-end workflow from syntax to numerics.

This paper introduces \texttt{Optyx}\footnote{Source code available at
\url{https://github.com/quantinuum-dev/optyx}.}, an open-source Python package that fills this gap. The package is an implementation of the dataflow programming framework for linear optical distributed quantum computing \cite{defelice2025dataflow}. Building on the categorical quantum mechanics ecosystem (\texttt{DisCoPy}, \texttt{PyZX}), \texttt{Optyx} offers a compositional domain-specific language in which gates, measurements, states, and quantum channels can be wired together into \emph{string diagrams}. These diagrams are compiled automatically into ZX/ZW fragments and subsequently into optimised tensor-network contractions using \texttt{Cotengra} and \texttt{Quimb}. The result is a simulator that (i) supports both qubit and discrete-variable photonic modes in a common formalism, (ii) transparently handles classical control, mid-circuit feedback, and probabilistic branching, and (iii) exposes differentiable backends for gradient-based design and calibration of hybrid experiments.

Beyond numerical simulation, we position \texttt{Optyx} as a \emph{prototyping environment} for networked quantum optics. Users may encapsulate sub-circuits as reusable components, assemble them into layered architectures connected by ideal or lossy channels, and deploy the same description to either an exact or approximate TN backend or (in the future) on hardware.

In the remainder of this article we: (i) give an overview of the software: basic syntax and evaluation methods (\S2); (ii) present the core building blocks provided by \texttt{Optyx} (\S3); (iii) discuss how to model noise and errors (\S4); (iv) describe conversion of circuits to and from other software (\S5); (v) detail the translation from diagrams to optimised tensor networks (\S6); and (vi) demonstrate the capabilities of the framework on three representative use cases: simulation of photonic observables with entangled states, distributed entanglement generation, and variational optimisation (\S7).
\section{Overview}
\label{sec:overview}
Our library provides a diagrammatic toolkit for prototyping and simulating hybrid quantum architectures. It exposes:
\begin{enumerate}
  \item a \textbf{functional front-end} (Pythonic, compositional),
  \item a compact \textbf{intermediate representation} for ZX/ZW generators with classical control, and
  \item pluggable \textbf{backends} (e.g., exact, tensor-network, permanent-based) for evaluation and \textbf{converters} for importing/exporting circuits from/to external tools.
\end{enumerate}

\paragraph{Basic syntax}
The circuits in \texttt{Optyx} (which we call \textit{diagrams} following the convention from the string diagram literature) are composed of generators representing quantum channels (CPTP maps). Because the
composition rules allow for mixing classical and quantum wires,
users can freely interleave them, routing classical measurement results into photonic or qubit feed-forward corrections, embedding qubits in photonic interferometers by using the dual-rail encoding where logical states are encoded in the single-photon subspace of two modes \cite{knill_scheme_2001}, or calling classical post-processing routines without leaving the
tensor-network formalism which provides the semantics for all circuits in \texttt{Optyx}.  Hybrid circuits are therefore \emph{first-class citizens} that compile, optimise, and execute through the very same tensor-network-based backend.

In \texttt{Optyx}, boxes represent maps, wiring boxes in series denotes sequential composition, and placing them side by side denotes the monoidal (parallel) product. Concretely, we follow the \texttt{DisCoPy} style: a \emph{diagram} is a typed compositional object. A map $f:A\rightarrow B$ can be composed in parallel (tensored) with $g:C\rightarrow D$ to obtain $f\otimes g:A\otimes C\rightarrow B\otimes D$, or composed in sequence when codomain and domain coincide; in code we write \verb|@| and \verb|>>| respectively. The same high-level diagram can later be interpreted as a tensor network, or a permanent computation. Four primitive system types appear throughout: \texttt{bit} for classical two-level data, \texttt{mode} for classical natural numbers (e.g. photon counts), \texttt{qubit} for two-dimensional quantum systems, and \texttt{qmode} for discrete-variable photonic modes. Qubit sub-diagrams are mostly expressed in the ZX-calculus, whose green (Z) and red (X) spiders form two interacting commutative Frobenius algebras \cite{coecke_interacting_2011}; the related ZW-calculus is convenient for infinite-dimensional discrete Hilbert spaces and features Z spiders, the binomial split/merge maps (the W nodes), and photon creations and selections \cite{light-matter-2023}. \texttt{Optyx} uses ZX/ZW diagrams to enable a conversion to tensor networks.






\begin{table*}[t]
\centering
\begin{tabular}{@{}ll@{}}
\toprule
\textbf{Module} & \textbf{Description} \\ \midrule
\texttt{optyx.channel} & Wire swaps; identity channels; discarding, measurement and preparation\\
\texttt{optyx.qubits}     & ZX gates; imported external circuits; qubit errors \\
\texttt{optyx.photonic}  & Linear-optical \& dual-rail operators; photon sources; detectors \\
\texttt{optyx.classical} & Boolean logic; $\mathbb{N}$ arithmetic; post-selection; control boxes \\ \bottomrule
\end{tabular}
\caption{The main sub-modules of \texttt{Optyx} and their functionality.}
\label{submodules_table}
\end{table*}

\paragraph{Function syntax} The example of teleportation (see Code~Listing~\ref{lst:teleportation-zx}) showcases  both sequential (\texttt{>>}) and parallel composition (\texttt{@}) to build a hybrid protocol. Alternatively, we can use the \textit{function syntax} instead of manual composition using \texttt{>>}. In this case, the generators act as functions defined on their domain wires. This approach is particularly useful when a diagram would otherwise require the user to manually insert \texttt{Swap} boxes to reorder wires. Since the wires are labelled using function syntax, the user can directly reference these labels as inputs to the appropriate boxes. The function syntax is similar to how the quantum programs are defined in \texttt{Guppy} \cite{koch2024guppy}. We provide an example of both the monoidal and function syntax in Code~Listing~\ref{lst:teleportation-zx}.

\paragraph{Evaluation with backends}
Numerical simulation proceeds by translating diagrams to tensor networks: each generator becomes a tensor, each wire an index, and diagram composition becomes index contraction. The difficulty of simulation is governed by the contraction order (path) and intermediate tensor sizes. \texttt{Optyx} compiles diagrams by (i) converting them to tensor networks, and (ii) handing the resulting TN to \texttt{Cotengra} for path optimisation and to \texttt{Quimb} for contraction (CPU/GPU, dense/sparse). TNs scale well when circuits have low treewidth and shallow depth; they fail sharply when intermediate tensors exceed memory.

For purely linear-optical amplitudes the standard alternative is permanent-based simulation: for an $m\times m$ interferometer $U$ acting on an $n$-photon Fock input $|\mathbf{s}\rangle$ (with $\sum_i s_i=n$), the amplitude for outcome $|\mathbf{t}\rangle$ is,
\begin{equation}
\langle\mathbf{t}|U|\mathbf{s}\rangle = \frac{\mathrm{Perm}(U_{\mathbf{t},\mathbf{s}})}{\sqrt{\prod_i s_i!\,\prod_j t_j!}},
\end{equation}
where $U_{\mathbf{t},\mathbf{s}}$ repeats rows/columns according to occupations. Computing the permanent is \#P-hard~\cite{valiant_complexity_1979}, with best-known exact algorithms scaling like $O(2^n n)$ (e.g., Ryser~\cite{ryser_combinatorial_1963}, Glynn~\cite{glynn_permanent_2010} algorithms). For approximate permanent calculation, Gurvits' algorithm gives an additive estimate of a permanent: for an \(n\times n\) matrix \(A\) it computes \(\operatorname{Per}(A)\) within \(\pm \varepsilon \|A\|^n\) in \(O(n^2/\varepsilon^2)\) time; because linear-optical output amplitudes are permanents of unitary submatrices, this yields efficient additive approximations to \emph{individual photonic outcome probabilities} \cite{Aaronson_gurvits_2014}. Toolkits such as \texttt{Perceval}~\cite{Heurtel2023percevalsoftware} provide permanent-based evaluators, SLOS~\cite{heurtel_strong_2023}, and Gurvits' which are advantageous when only a few amplitudes are needed or when $n$ is small. Consequently, permanent-based methods excel for typical low-width passive linear-optical circuits, whereas adding active control and feed-forward, non-unitarities, or hybrid qubit--photon models quickly becomes cumbersome for permanent-based approaches (with these setups the number of permanents to calculate can scale exponentially). By contrast, the tensor-network methods we describe in this article natively handle these features. \texttt{Optyx} exposes both TN and permanent-based backends.

Diagrams can be evaluated using the \texttt{Quimb} backend (exact or approximate evaluation), the \texttt{Perceval} backend\footnote{Note that currently only linear optical channels can be evaluated using this method.} (with permanent-based algorithms), or the \texttt{DisCoPy} backend. By default, evaluation is performed with \texttt{Quimb} exact contraction: \texttt{result = circ.eval()}. The \texttt{result} object provides ways to obtain amplitudes of the resulting state, probability distribution, or a density matrix.

\paragraph{Example: qubit teleportation}

\begin{figure}[H]
  \centering
  \includegraphics[width=0.7\linewidth]{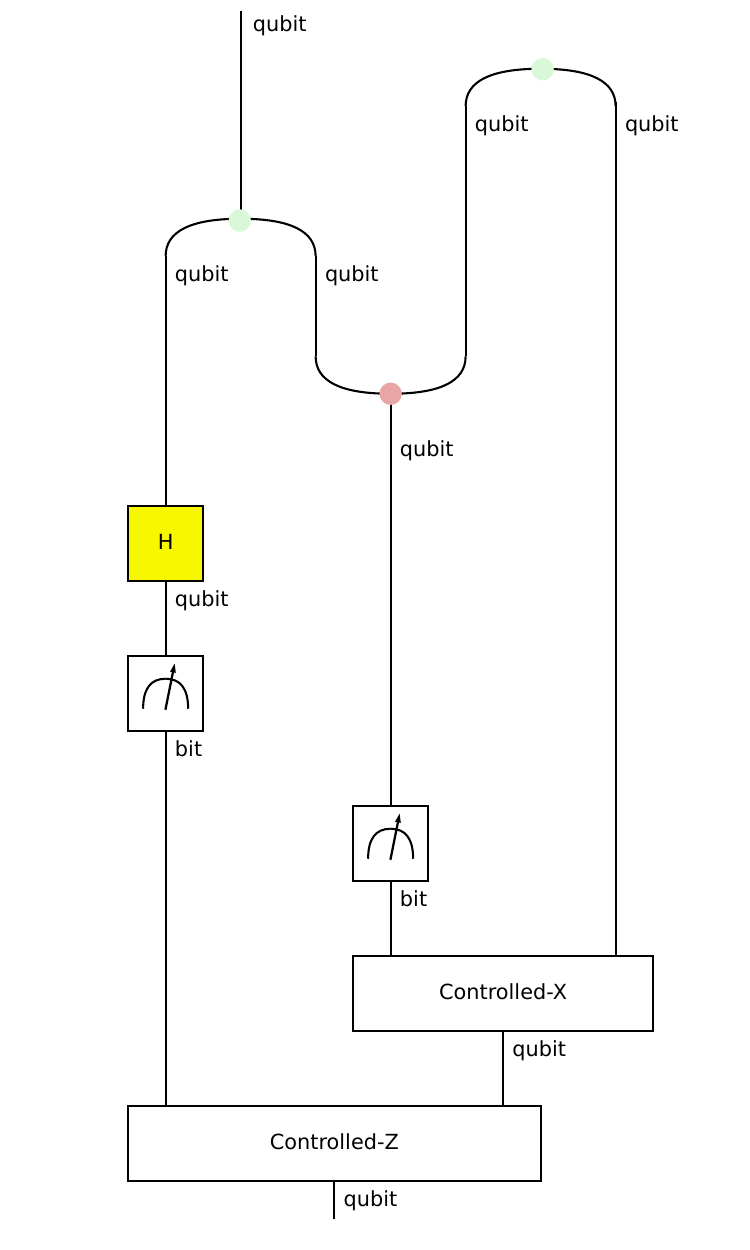}
  \caption{Diagram of the teleportation protocol in ZX-calculus. We discuss quantum teleportation using fusion measurements in the Appendix~\ref{sec:teleportation-fusion}.}
  \label{fig:teleportation-protocol-zx}
\end{figure}

Quantum teleportation uses classical and quantum data (it is therefore a \emph{classical-quantum ZX diagram}). In \texttt{Optyx}, we can make use of controlled boxes from \texttt{optyx.classical} to correct the state of the qubit depending on the measurement result (see Figure~\ref{fig:teleportation-protocol-zx}).

\begin{lstlisting}[language=Python,
  caption={Teleportation protocol with classical control using both monoidal and function syntax.},
  label={lst:teleportation-zx}]
@Channel.from_callable(
  dom=qubit @ qubit, cod=qubit @ qubit
)
def cnot(a, b):
  c, d = Z(1, 2)(a)
  Scalar(2 ** 0.5)()
  return c, X(2, 1)(d, b)

bell = Scalar(0.5 ** 0.5) @ Z(0, 2)

@Channel.from_callable(
  dom=qubit, cod=qubit
)
def teleportation(c):
  a, b = bell()
  cc, aa = cnot(c, a)
  c_ = Measure(1)(H()(cc))
  a_ = Measure(1)(aa)
  bb = CtrlX(a_, b)
  return CtrlZ(c_, bb)

teleportation_monoidal_syntax = (
  qubit @ bell >>
  cnot @ qubit >>
  H() @ qubit ** 2 >>
  Measure(1) @ Measure(1) @ qubit >>
  bit @ CtrlX >>
  CtrlZ
)
\end{lstlisting}

In the following sections, we discuss the main building blocks of \texttt{Optyx} in more detail: the \texttt{qubits} and \texttt{photonic} modules. Many examples make use of the \texttt{classical} module to perform classical operations on the results of quantum measurements, such as post-selection, copying, or discarding bits.

\section{Building blocks}\label{sec:building-blocks}
We have three kinds of generators: \texttt{qubits} generators, \texttt{photonic} generators, and \texttt{classical} generators (see Table~\ref{submodules_table}). Intuitively, they represent gates and transformations on appropriate types.
The generators make use of the following types:
\texttt{bit} and \texttt{mode} for classical information (bit and natural number arithmetic respectively), and
\texttt{qubit} and \texttt{qmode} for quantum degrees of freedom (qubit and photonic (Fock space) types). At the end of this section (see Subsection~\ref{sec:stream-processes}), we also discuss the stream processes which allow us to build synchronous circuits with feedback loops and delays and execute them for a specified number of time-steps.

\subsection{General channels}
There are generators in \texttt{Optyx} which belong to all the classes (\texttt{qubits}, \texttt{photonic}, and \texttt{classical}) and their specific use depends on the types supplied.

\paragraph{\texttt{Channel} class} The superclass of all generators in \texttt{Optyx}. It can be used to define custom channels (maps) in \texttt{Optyx}. It is achieved by supplying a Kraus map to a \texttt{Channel} (see Appendix~\ref{sec:kraus_maps}) together with the types for the domain and codomain.

\paragraph{Classical-quantum generators} These are the generators which interface between classical types (\texttt{bit} and \texttt{mode}) and quantum types (\texttt{qubit} and \texttt{qmode}):

\begin{itemize}
  \item \texttt{Discard(type)}: discards a wire of type \texttt{type}. For example, we can use this to trace out qubit registers.
  \item \texttt{Measure(type)}: by measuring a quantum type, we obtain a corresponding classical register ($\text{\texttt{qubit}}^{\otimes m} \rightarrow \text{\texttt{bit}}^{\otimes m}$ and $\text{\texttt{qmode}}^{\otimes m} \rightarrow \text{\texttt{mode}}^{\otimes m}$).
  \item \texttt{Encode(type)}: the reverse of \texttt{Measure(type)} - we can use this to encode classical data in quantum registers (state preparation).
\end{itemize}

We also have other generators, which do not fall into the qubit, photonic or classical categories and require data types to be specified. These are \texttt{Id}, \texttt{Swap}, \texttt{Scalar}, \texttt{Spider}, and \texttt{Copy}. For example \texttt{Id(mode**2 @ qubit)} is an identity operator on two modes and one qubit.

\subsection{Qubit primitives}
The sub-module \texttt{optyx.qubits} collects every generator that acts on
qubit registers:
ZX-calculus generators, wrappers around circuit
descriptions from external sources, basis preparations and measurements, and a handful of
noise models (see Table~\ref{tab:qubit_primitives}). As other external packages (like \texttt{{tket}} or \texttt{{Graphix}}) offer well-established circuit or measurement pattern descriptions, we provide a way to wrap these objects into the \texttt{Optyx} framework, while also maintaining a basic set of generators that can be used to construct circuits from scratch directly in \texttt{Optyx}.

\begin{table*}[t]
  \centering
  \caption{Summary of available qubit primitives}
  \begin{tabular}{@{}lp{5.5cm}p{7.5cm}@{}}
    \toprule
    \textbf{Category} & \textbf{Primitives} & \textbf{Description} \\
    \midrule
    Circuits     & \texttt{Circuit}, \texttt{QubitChannel} & Quantum circuits and channels operating on qubits. \\
    ZX           & \texttt{Z}, \texttt{X}, \texttt{H}, \texttt{Scalar}, \texttt{Bra}, \texttt{Ket} & Core ZX-calculus generators (spiders, Hadamard, scalar). \\
    Errors       & \texttt{BitFlipError}, \texttt{DephasingError} & Stochastic error channels acting on qubits. \\
    \bottomrule
  \end{tabular}
  \label{tab:qubit_primitives}
\end{table*}

The main way of defining qubit maps in \texttt{Optyx} is using ZX diagrams. In particular, we are interested in using them to represent, for example, graph states for measurement-based quantum computing (MBQC) \cite{rausendorf_measurement-based_2009} or fusion-based computing \cite{browne_resource-efficient_2005}. We include the usual generators of ZX-calculus such as \texttt{Z}, \texttt{X}, and \texttt{H}, which can be imported from both \texttt{optyx.classical} and \texttt{optyx.qubits}. The classical versions act on \texttt{bit} types, while the quantum versions act on \texttt{qubit} types. For example, \texttt{X(1, 1, 0.5)} imported from \texttt{optyx.classical} represents a classical NOT gate, whereas the same gate imported from \texttt{optyx.qubits} corresponds to a Pauli-X gate acting on a qubit.

Since we are dealing with mixed channels, we also include generators that interface between classical and quantum systems: \texttt{Encode} (encoding a classical \texttt{bit} into a \texttt{qubit}), \texttt{Measure} (measuring a qubit in the computational basis to obtain a \texttt{bit}), and \texttt{Discard} (tracing out a qubit or deleting a bit). Using these, one can build \emph{classical-quantum ZX diagrams} that mix quantum ($\text{\texttt{qubit}}^{\otimes m} \rightarrow \text{\texttt{qubit}}^{\otimes n}$) and classical ($\text{\texttt{bit}}^{\otimes m} \rightarrow \text{\texttt{bit}}^{\otimes n}$) components within a single diagram.












\subsection{Photonic primitives}

\begin{table*}[t]
  \centering
  \caption{Summary of available photonic primitives}
  \begin{tabular}{@{}lp{5cm}p{7cm}@{}}
    \toprule
    \textbf{Category} & \textbf{Primitives} & \textbf{Description}\\
    \midrule
    Linear optical gates
      & \texttt{Gate}, \texttt{Phase}, \texttt{BBS}, \texttt{TBS}, \texttt{MZI}, \texttt{ansatz}
      & Unitary transformations built from beam splitters, phase shifters, and general interferometers.\\
    Dual-rail operators
      & \texttt{DualRail}, \texttt{HadamardBS}, \texttt{PhaseShiftDR}, \texttt{ZMeasurementDR}, \texttt{XMeasurementDR}, \texttt{FusionTypeI}, \texttt{FusionTypeII}
      & Gates and measurements for qubits encoded in dual-rail photonic modes.\\
    Measurements
      & \texttt{PhotonThresholdMeasurement}, \texttt{NumberResolvingMeasurement}
      & Operations that measure photon number.\\
    States
      & \texttt{Create}
      & Preparation of bosonic product states.\\
    Other
      & \texttt{NumOp}, \texttt{PhotonLoss}
      & Miscellaneous operators.\\
    \bottomrule
  \end{tabular}
  \label{photonic_primitives_table}
\end{table*}

Discrete-variable modes are implemented in \texttt{optyx.photonic}.
The module organises its generators into five categories summarised in
Table~\ref{photonic_primitives_table}: measurements, linear-optical
gates, dual-rail operators, states, and utility channels.
Below each category is described in more detail.

\paragraph{Linear-optical gates}
All passive linear optical gates support \verb|.dagger()| and \verb|.conjugate()|, accept symbolic
parameters, and are differentiable.
\noindent
\begin{itemize}
  \item \texttt{BBS(bias)} -- biased beam-splitter, balanced when $\mathtt{bias}=0$; reproduces
        the Hong-Ou-Mandel dip \cite{hong_measurement_1987}.
  \item \texttt{TBS(phi)} -- continuously tunable beam-splitter with mixing
        angle \(\theta\).
  \item \texttt{MZI(psi,phi)} -- two-phase Mach-Zehnder interferometer.
  \item \texttt{Phase(psi)} -- single-mode phase shift
\end{itemize}

\paragraph{Measurements}
In \texttt{Optyx}, we provide two types of photonic measurements:
\begin{itemize}
  \item \texttt{PhotonThresholdMeasurement} -- a click/no-click detector
        that returns a single classical bit (1 if at least one photon is
        present).
  \item \texttt{NumberResolvingMeasurement(n)} -- resolves the full
        photon-count pattern on \(n\) modes and outputs
        \texttt{mode}$^{\otimes n}$.
\end{itemize}

\paragraph{Photon sources - dual-rail and Fock states}
The \texttt{Create} generator prepares Fock states on $n$ modes. It accepts a list of occupations and an optional list of internal states (to model photon distinguishability, see Section~\ref{sec:noise-and-errors}). Likewise, \texttt{DualRail} encodes a qubit into two photonic modes using a single photon. It also optionally accepts a list of internal states (internal degrees of freedom for circuits with varying degrees of photon distinguishability, see Section~\ref{sec:noise-and-errors}).

\subsection{Classical layer and hybrid control}

\begin{table*}[t]
  \centering
  \caption{Summary of available \emph{classical} primitives}
  \begin{tabular}{@{}lp{5cm}p{7cm}@{}}
    \toprule
    \textbf{Category} & \textbf{Primitives} & \textbf{Description}\\
    \midrule
    Logic gates
      & \texttt{Not}, \texttt{Xor}, \texttt{And}, \texttt{Or}, \texttt{Z}, \texttt{X}
      & Boolean operations and ZX spiders acting on classical bits.\\
    Arithmetic on modes
      & \texttt{Add}, \texttt{Sub}, \texttt{Multiply}, \texttt{Divide},
        \texttt{Mod2}
      & Integer arithmetic and data-movement primitives on natural-number modes.\\
    Control \& feedforward
      & \texttt{BitControlledGate}, \texttt{BitControlledPhaseShift},
        \texttt{CtrlX}, \texttt{CtrlZ},
        \texttt{ClassicalFunction}, \texttt{BinaryMatrix}
      & Classical control of quantum gates and generic functions/matrix maps on bit or mode registers.\\
    \bottomrule
  \end{tabular}
  \label{classical_primitives_table}
\end{table*}

The \texttt{optyx.classical} module supplies logic, arithmetic on
natural-number modes, and control boxes that condition quantum
sub-circuits on classical data (see Table~\ref{classical_primitives_table}). The generators are organised in two categories: primitives for classical computations (obtainable from explicit generators - logical gates and arithmetic on modes, \texttt{Python} functions\footnote{
  To obtain a tensor for a given \texttt{Python} function, \texttt{Optyx} evaluates the function on all possible inputs for given input dimensions and constructs a tensor with the corresponding output values.
} and Boolean matrices), and primitives acting on quantum types (\texttt{qubit} and \texttt{qmode}) that condition quantum gates on classical data (e.g. \texttt{BitControlledGate}). Hybrid control is obtained with \texttt{BitControlledGate} and \texttt{BitControlledPhaseShift}.

\medskip
\noindent
Taken together, these generators let users script algorithms, graph states, realistic error channels, and
hybrid qubit-photon experiments - \emph{all} within a single
tensor-network semantics that compiles to either \texttt{Quimb} contractions or
permanent-based linear-optical paths as required.

\subsection{Feedback and delay}
\label{sec:stream-processes}
Modern optical setups with sources, routers, and delay lines require us to reason
about how systems evolve over many time-steps with feedback. Stream processes
are diagrams for such time-extended behaviours: they describe (potentially
infinite) sequences of inputs and outputs and provide a language to analyse and
transform realistic optical circuits. For a formal definition of streams, see
Ref.~\cite{defelice2025dataflow}, Section~4.

Informally, a stream process from an input stream $X$ to an output stream $Y$
is a circuit executed repeatedly in discrete time-steps, together with an
internal memory carried from one step to the next. At time-step~$0$ the process
takes $X_0$ and an initial memory $M_0$, produces $Y_0$, and updates the memory
to $M_1$. From time-step~$1$ onwards, the same pattern repeats with
$(X_i, M_i)$ producing $(Y_i, M_{i+1})$. Unfolding this recursion, a stream is
determined by its family of one-step maps
\[
  f_i : M_i \otimes X_i \to M_{i+1} \otimes Y_i \qquad (i = 0,1,2,\dots),
\]
where $M_i$ is the memory linking time-steps.

In this setting, feedback is the operation that creates and updates this memory:
it takes a process with an extra input/output system $S$ and connects those
wires in a loop, so that the value of $S$ produced at one time-step is stored
and later fed back as an input. Physically, this corresponds to sending part of
the signal through a delay line or quantum memory and re-injecting it into the
setup.

A basic example built from feedback is a delay. To implement a delay we use a stream of length $1$ on $X$ that takes the current value at time-step $i$, stores it in memory,
and outputs instead the value stored at time-step $i-1$. Diagrammatically, this
is obtained from a suitable feedback of a swap:
\[
  \scalebox{0.7}{\begin{tikzpicture}
	\begin{pgfonlayer}{nodelayer}
		\node [style=none] (33) at (-3.5, 2.25) {$=$};
		\node [style=none] (35) at (-3.5, -3.75) {$=$};
		\node [style=none] (82) at (-8.75, 1.5) {};
		\node [style=none] (83) at (-4.25, 1.5) {};
		\node [style=none] (90) at (-7.75, 2.5) {};
		\node [style=none] (91) at (-7.75, 3.5) {};
		\node [style=none] (92) at (-5.25, 2.5) {};
		\node [style=none] (93) at (-5.25, 3.5) {};
		\node [style=none] (95) at (-7.5, 1.5) {};
		\node [style=none] (96) at (-5.5, 1.5) {};
		\node [style=none] (97) at (-7.5, 2.5) {};
		\node [style=none] (98) at (-5.5, 2.5) {};
		\node [style=none] (100) at (2.25, 1.5) {};
		\node [style=none] (101) at (6.75, 1.5) {};
		\node [style=none] (102) at (1.3, 1.5) {};
		\node [style=none] (103) at (3.25, 2.5) {};
		\node [style=none] (104) at (3.25, 3.5) {};
		\node [style=none] (105) at (5.75, 2.5) {};
		\node [style=none] (106) at (5.75, 3.5) {};
		\node [style=none] (108) at (3.5, 1.5) {};
		\node [style=none] (109) at (5.5, 1.5) {};
		\node [style=none] (110) at (3.5, 2.5) {};
		\node [style=none] (111) at (5.5, 2.5) {};
		\node [style=none] (113) at (-2.5, 3) {};
		\node [style=none] (117) at (-2, 3) {};
		\node [style=none] (120) at (0.5, 1.5) {};
		\node [style=none] (121) at (1.75, 4.5) {};
		\node [style=none] (122) at (1.75, 0.5) {};
		\node [style=none] (123) at (-0.75, 4.25) {$t_0$};
		\node [style=sLabel] (156) at (-8.5, 0.875) {$X$};
		\node [style=sLabel] (157) at (-4.5, 0.875) {$X$};
		\node [style=sLabel] (162) at (2.5, 0.875) {$X$};
		\node [style=sLabel] (163) at (6.5, 0.875) {$X$};
		\node [style=none] (164) at (-8.225, 3) {};
		\node [style=black node] (165) at (-9, 3) {};
		\node [style=none] (166) at (-9, 3.5) {$0$};
		\node [style=black node] (167) at (-2.5, 3) {};
		\node [style=none] (168) at (-2.5, 3.5) {$0$};
		\node [style=none] (172) at (2.8, 3) {};
		\node [style=none] (173) at (-2.5, 1.5) {};
		\node [style=none] (174) at (-2, 1.5) {};
		\node [style=none] (175) at (0.5, 3) {};
		\node [style=none] (178) at (1.05, -4.5) {};
		\node [style=none] (187) at (-2.5, -3) {};
		\node [style=none] (188) at (-2, -3) {};
		\node [style=none] (189) at (0.5, -4.5) {};
		\node [style=none] (190) at (1.5, -1.5) {};
		\node [style=none] (191) at (1.5, -5.5) {};
		\node [style=none] (192) at (-0.75, -1.75) {$t_0$};
		\node [style=black node] (195) at (-2.5, -3) {};
		\node [style=none] (196) at (-2.5, -2.5) {$0$};
		\node [style=none] (197) at (2, -3) {};
		\node [style=none] (198) at (-2.5, -4.5) {};
		\node [style=none] (199) at (-2, -4.5) {};
		\node [style=none] (200) at (0.5, -3) {};
		\node [style=none] (201) at (6.5, -4.5) {};
		\node [style=none] (202) at (11, -4.5) {};
		\node [style=none] (203) at (5.55, -4.5) {};
		\node [style=none] (204) at (7.5, -3.5) {};
		\node [style=none] (205) at (7.5, -2.5) {};
		\node [style=none] (206) at (10, -3.5) {};
		\node [style=none] (207) at (10, -2.5) {};
		\node [style=none] (208) at (7.75, -4.5) {};
		\node [style=none] (209) at (9.75, -4.5) {};
		\node [style=none] (210) at (7.75, -3.5) {};
		\node [style=none] (211) at (9.75, -3.5) {};
		\node [style=none] (212) at (2, -3) {};
		\node [style=none] (213) at (2.5, -3) {};
		\node [style=none] (214) at (5, -4.5) {};
		\node [style=none] (215) at (6, -1.5) {};
		\node [style=none] (216) at (6, -5.5) {};
		\node [style=none] (217) at (3.75, -1.75) {$t_1$};
		\node [style=sLabel] (218) at (6.75, -5.125) {$X$};
		\node [style=sLabel] (219) at (10.75, -5.125) {$X$};
		\node [style=none] (222) at (7.05, -3) {};
		\node [style=none] (223) at (2, -4.5) {};
		\node [style=none] (224) at (2.5, -4.5) {};
		\node [style=none] (225) at (5, -3) {};
	\end{pgfonlayer}
	\begin{pgfonlayer}{edgelayer}
		\draw [style=NLO, bend left=270, looseness=1.75] (92.center) to (93.center);
		\draw [style=NLO, bend left=270, looseness=1.50] (91.center) to (90.center);
		\draw [style=NLO] (83.center) to (96.center);
		\draw [style=NLO] (82.center) to (95.center);
		\draw [style=NLO] (92.center) to (98.center);
		\draw [style=NLO] (97.center) to (90.center);
		\draw [style=NLO, in=180, out=0] (95.center) to (98.center);
		\draw [style=NLO, in=180, out=0] (97.center) to (96.center);
		\draw [style=NLO, bend left=270, looseness=1.75] (105.center) to (106.center);
		\draw [style=NLO, bend left=270, looseness=1.50] (104.center) to (103.center);
		\draw [style=NLO] (101.center) to (109.center);
		\draw [style=NLO] (100.center) to (108.center);
		\draw [style=NLO] (105.center) to (111.center);
		\draw [style=NLO] (110.center) to (103.center);
		\draw [style=NLO, in=180, out=0] (108.center) to (111.center);
		\draw [style=NLO, in=180, out=0] (110.center) to (109.center);
		\draw [style=LOWire] (113.center) to (117.center);
		\draw [style=LOWire, in=180, out=0] (117.center) to (120.center);
		\draw [style=LOWire] (120.center) to (102.center);
		\draw [style=dashedE] (121.center) to (122.center);
		\draw [style=LOStream] (93.center) to (91.center);
		\draw [style=LOStream] (106.center) to (104.center);
		\draw (165) to (164.center);
		\draw [style=LOWire] (173.center) to (174.center);
		\draw [style=LOWire, in=180, out=0] (174.center) to (175.center);
		\draw [style=LOWire] (175.center) to (172.center);
		\draw [style=LOWire] (187.center) to (188.center);
		\draw [style=LOWire, in=180, out=0] (188.center) to (189.center);
		\draw [style=LOWire] (189.center) to (178.center);
		\draw [style=dashedE] (190.center) to (191.center);
		\draw [style=LOWire] (198.center) to (199.center);
		\draw [style=LOWire, in=180, out=0] (199.center) to (200.center);
		\draw [style=LOWire] (200.center) to (197.center);
		\draw [style=NLO, bend left=270, looseness=1.75] (206.center) to (207.center);
		\draw [style=NLO, bend left=270, looseness=1.50] (205.center) to (204.center);
		\draw [style=NLO] (202.center) to (209.center);
		\draw [style=NLO] (201.center) to (208.center);
		\draw [style=NLO] (206.center) to (211.center);
		\draw [style=NLO] (210.center) to (204.center);
		\draw [style=NLO, in=180, out=0] (208.center) to (211.center);
		\draw [style=NLO, in=180, out=0] (210.center) to (209.center);
		\draw [style=LOWire] (212.center) to (213.center);
		\draw [style=LOWire, in=180, out=0] (213.center) to (214.center);
		\draw [style=LOWire] (214.center) to (203.center);
		\draw [style=dashedE] (215.center) to (216.center);
		\draw [style=LOStream] (207.center) to (205.center);
		\draw [style=LOWire] (223.center) to (224.center);
		\draw [style=LOWire, in=180, out=0] (224.center) to (225.center);
		\draw [style=LOWire] (225.center) to (222.center);
	\end{pgfonlayer}
\end{tikzpicture}
}
\]
A delay of length $d$ is given by composing $d$ unit delays, so that each value
remains in memory for $d$ time-steps before being output. The delay can be
implemented in \texttt{Optyx} as in Code~Listing~\ref{lst:stream-delay}.

\begin{lstlisting}[language=Python,
  caption={Delay on a \texttt{qmode} with an empty initial state.},
  label={lst:stream-delay}]
from optyx import Diagram, qmode
from optyx.photonic import Create

Diagram.delay(
  ty=qmode,
  initial_state=Create(0)
)
\end{lstlisting}

Other processes can also be modelled with feedback -- consider unrolling of a CNOT ladder:
\[
  \begin{tikzpicture}
	\begin{pgfonlayer}{nodelayer}
		\node [style=Z dot] (0) at (-0.75, 1) {};
		\node [style=X dot] (1) at (-0.75, 0) {};
		\node [style=none] (2) at (0.5, 1) {};
		\node [style=none] (3) at (0.5, 0) {};
		\node [style=none] (4) at (-1.75, 1) {};
		\node [style=none] (5) at (-3, 0) {};
		\node [style=none] (6) at (-1.75, 2) {};
		\node [style=none] (7) at (0.5, 2) {};
		\node [style=none] (8) at (1.25, 0) {};
		\node [style=none] (9) at (-2.225, 1.5) {};
		\node [style=Z dot] (10) at (-3, 1.5) {};
		\node [style=none] (11) at (3.75, 0.5) {$\mapsto$};
		\node [style=Z dot] (12) at (7.5, 2.5) {};
		\node [style=Z dot] (13) at (8.5, 1.5) {};
		\node [style=Z dot] (14) at (9.5, 0.5) {};
		\node [style=X dot] (16) at (7.5, 1.5) {};
		\node [style=X dot] (17) at (8.5, 0.5) {};
		\node [style=X dot] (18) at (9.5, -0.5) {};
		\node [style=Z dot] (19) at (6.5, 2.5) {};
		\node [style=none] (20) at (6.5, 1.5) {};
		\node [style=none] (21) at (6.5, 0.5) {};
		\node [style=none] (22) at (6.5, -0.5) {};
		\node [style=none] (23) at (11.5, 2.5) {};
		\node [style=none] (24) at (11.5, 1.5) {};
		\node [style=none] (25) at (11.5, 0.5) {};
		\node [style=none] (26) at (11.5, -0.5) {};
		\node [style=none] (27) at (3.75, 1.5) {$\mathtt{unroll}_3$};
		\node [style=Z dot] (28) at (10.5, -0.5) {};
		\node [style=X dot] (29) at (10.5, -1.5) {};
		\node [style=none] (30) at (6.5, -1.5) {};
		\node [style=none] (31) at (11.5, -1.5) {};
	\end{pgfonlayer}
	\begin{pgfonlayer}{edgelayer}
		\draw [style=N, in=-180, out=0, looseness=0.75] (1) to (2.center);
		\draw [style=N, in=-180, out=0] (0) to (3.center);
		\draw [style=N] (1) to (0);
		\draw [style=N, bend right=90, looseness=2.25] (2.center) to (7.center);
		\draw [style=N, bend right=270, looseness=1.50] (4.center) to (6.center);
		\draw [style=N] (4.center) to (0);
		\draw [style=N] (5.center) to (1);
		\draw [style=NStream] (7.center) to (6.center);
		\draw [style=N] (3.center) to (8.center);
		\draw (10) to (9.center);
		\draw (19) to (12);
		\draw (12) to (23.center);
		\draw (20.center) to (16);
		\draw (16) to (13);
		\draw (13) to (24.center);
		\draw (21.center) to (17);
		\draw (17) to (14);
		\draw (14) to (25.center);
		\draw (22.center) to (18);
		\draw (18) to (26.center);
		\draw (12) to (16);
		\draw (13) to (17);
		\draw (14) to (18);
		\draw (30.center) to (29);
		\draw (29) to (28);
		\draw (29) to (31.center);
	\end{pgfonlayer}
\end{tikzpicture}

\]

We can implement this exact stream in \texttt{Optyx} like in Code~Listing~\ref{lst:stream-cnot}. Here \texttt{diagram} defines the one-step body of the process; the call \texttt{diagram.feedback(dom=..., cod=..., mem=..., initial\_state=...)} applies the feedback construction, turning this body into a stream by introducing a single-qubit memory wire that is threaded between time-steps; and the final call to \texttt{.unroll(3)} materialises the first three time-steps of this infinite stream as a finite diagram.

\begin{lstlisting}[language=Python,
  caption={CNOT ladder stream process.},
  label={lst:stream-cnot}]
diagram = (
  X(1, 2) @ qubit >>
  qubit @ Z(2, 1) >>
  Swap(qubit, qubit)
)

diagram.feedback(
  dom=qubit,
  cod=qubit,
  mem=qubit,
  initial_state=Z(0, 1)
).unroll(3)

\end{lstlisting}

\section{Noise and errors}
\label{sec:noise-and-errors}
\paragraph{Photon distinguishability}
Nanosecond timing slips, energy shifts, or polarisation changes let two photons be told apart. This source of errors is called photon distinguishability. In photonic quantum computers this spoils the bosonic interference that most gates need, boosting mis-fires and error rates. Keeping photons indistinguishable is therefore as vital as low loss and high detector efficiency for scalable photonic processors. We model distinguishability in \texttt{Optyx} (see the example Code~Listing~\ref{lst:hom-dip-distinguishable}) by assigning \emph{internal states} to photon sources (in \texttt{Create} and \texttt{DualRail} instances). The pairwise inner products between internal states from different sources define the $S$ matrix of Ref.~\cite{menssen2017distinguishability}.

\paragraph{Photon loss} It is the disappearance of a photon through absorption, scattering, imperfect coupling, or detector inefficiency before it can participate in its intended operation. Because each photonic qubit exists in just one photon (in dual-rail encoding), losing that photon erases the quantum state entirely, so circuit success probabilities plummet as systems grow and errors accumulate. As photon-loss is the leading source of error in current photonic systems, it is important to be able to model it. In \texttt{Optyx}, we provide the \texttt{PhotonLoss} generator to model photon loss (see the example Code~Listing~\ref{lst:hom-dip-loss}). It accepts a transmission probability $p\in[0,1]$ and acts on \texttt{qmode} types. We are therefore able to model \emph{non-uniform} photon loss \cite{Brod2020classicalsimulation} by placing \texttt{PhotonLoss} boxes with different $p$ values in different locations.

Together, these primitives allow users to script anything from a single
Hong-Ou-Mandel benchmark to a loss-tolerant dual-rail cluster-state
factory, without leaving the tensor-network backend that powers \texttt{Optyx}.

\paragraph{Qubit errors}
We model single-qubit noise with two \texttt{qubit} generators. Two standard choices are the \emph{bit-flip} channel and the \emph{dephasing} (phase-flip) channel. For error probability $p\in[0,1]$, they act on a qubit state $\rho$ as
\[
\mathcal{E}_{X,p}(\rho)=(1-p)\,\rho + p\,X\rho X,
\]
\[
\mathcal{E}_{Z,p}(\rho)=(1-p)\,\rho + p\,Z\rho Z.
\]
In \texttt{Optyx}, these are provided as \texttt{BitFlipError(p)} and \texttt{DephasingError(p)} which act on the \texttt{qubit} type. They are used by inserting them \emph{where the noise occurs}: after a gate to model control errors, before a measurement to model readout noise, or around a subroutine to model a noisy block.

Other (custom) noise channels can be defined by a user by defining their own channel (see Appendix~\ref{sec:kraus_maps} for an explanation of how to define a custom channel).

\paragraph{Example: the Hong-Ou-Mandel effect} The effect arises when two indistinguishable photons strike opposite inputs of a 50:50 beamsplitter at the same moment; quantum interference drives them out through the same port, wiping out coincidence counts. The depth of this ``HOM dip'' measures indistinguishability, so a reduced dip flags timing, spectral, or polarisation mismatches that would lower photonic-gate fidelity. Let us consider the following example of a HOM dip with perfectly indistinguishable photons in Code~Listing~\ref{lst:hom-dip}. The probability of detecting one photon in each output mode is 0.0, while the probability of detecting both photons in one of the output modes is 0.5, which is the expected value for a perfectly indistinguishable pair of photons.

\begin{lstlisting}[language=Python,
  caption={Hong-Ou-Mandel effect with two indistinguishable photons.},
  label={lst:hom-dip}]
from optyx.photonic import BS, Create

beam_splitter = BS

HOM = (
    Create(1) @ Create(1) >>
    beam_splitter
)

assert np.allclose(
  HOM.eval().prob_dist()[1, 1],  0.0
)
\end{lstlisting}

In Code~Listing~\ref{lst:hom-dip-distinguishable}, the two internal states are random array-like vectors of unit norm: \texttt{s\_1} and \texttt{s\_2}. \texttt{Create} can accept internal states of photons as an argument. \texttt{Channel.inflate} is used to indicate evaluation of the channel taking into account the internal states. The probability of detecting one photon in each output mode should be $\frac{1}{2} - \frac{1}{2} \braket{s_1 | s_2}$.

\begin{lstlisting}[language=Python,
  caption={Hong-Ou-Mandel effect with two distinguishable photons.},
  label={lst:hom-dip-distinguishable}]
import numpy as np
from optyx.photonic import BS

s_1 = [1, 0]
s_2 = [np.sqrt(0.9), np.sqrt(0.1)]

create = Create(1, 1, internal_states=(
    s_1, s_2))

distinguishable_HOM = (
  create >> BS >>
  NumberResolvingMeasurement(2)
)
result = distinguishable_HOM.inflate(
    len(s_1)).eval().prob_dist()

theoretical_result = 0.5 - 0.5 * np.abs(
    np.array(s_1).dot(
        np.array(s_2).conjugate()))**2
assert np.isclose(result[(1, 1)], theoretical_result, 3)
\end{lstlisting}

With photon loss, the HOM effect does not hold anymore. We can actually observe one photon in one output mode with a non-zero probability. In Code~Listing~\ref{lst:hom-dip-loss} we model photon loss using the \texttt{PhotonLoss} generator. If we lose one of the two photons with probability $0.2$, then the probability of detecting one photon overall is also $0.2$.

\begin{lstlisting}[language=Python,
  caption={An example demonstrating photon loss. If we create two photons in two modes and lose one of them with probability $0.2$, then the probability of detecting one photon overall is $0.2$.},
  label={lst:hom-dip-loss}]
lossy_HOM = (
  Create(1, 1) >>
  PhotonLoss(0.8) @ Id(1) >>
  BS >>
  NumberResolvingMeasurement(2) >>
  Add(2)
)
assert np.isclose(
  lossy_HOM.eval().prob_dist()[(1,)], 0.2
)
\end{lstlisting}

\section{External interfaces}
\paragraph{Converting external qubit circuits}
A complete circuit or measurement pattern created in \texttt{tket}, \texttt{PyZX}, \texttt{Graphix}, or \texttt{DisCoPy} becomes a single channel with one line of code which can then be composed with other \texttt{Optyx} diagrams and evaluated using \texttt{Optyx}'s backends.

Any ZX diagram in \texttt{PyZX} can be translated into a \texttt{diagram.Diagram} using the method \texttt{from\_pyzx()} (this results only in pure diagrams). Quantum circuits defined in \texttt{tket}, or \texttt{DisCoPy} can also be converted into \texttt{Optyx} circuits (both pure and mixed).

A measurement pattern from \texttt{Graphix} \cite{Sunami:2022tub} can also be converted into an \texttt{Optyx} circuit. For example, we can define a circuit in \texttt{Graphix}, transpile it into a measurement pattern and then convert its underlying graph into an \texttt{Optyx} circuit.

\paragraph{Converting external photonic circuits} We can convert photonic circuits from \texttt{Perceval}~\cite{Heurtel2023percevalsoftware} into \texttt{Optyx}; for example, one can define a teleportation circuit with feed-forward in \texttt{Perceval} and then translate it into an equivalent \texttt{Optyx} circuit. While the converter does not support every \texttt{Perceval} feature (e.g. time delays), we can convert any circuit or processor that can be represented as a linear-optical circuit with feed-forward, including measurements, heralding, and post-selection.

\section{Construction and evaluation of tensor networks}\label{sec:construction-of-tensor-networks}
In \texttt{Optyx}, diagrams are the level of syntax, while tensor networks provide the operational semantics. The overall workflow is as follows: a user builds a diagram; a functorial translation turns that diagram into a tensor network; an optimiser finds an inexpensive contraction order; and an execution backend carries out the contractions numerically.

\subsection{From a diagram to a tensor network}

\begin{figure}[H]
  \centering
  \includegraphics[width=1.0\linewidth]{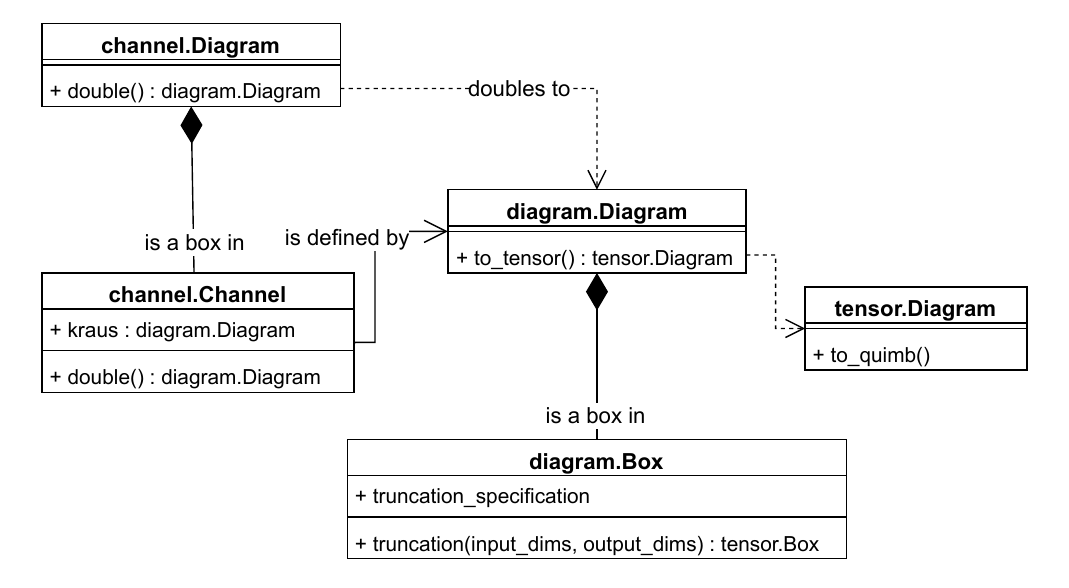}
  \caption{A UML class diagram \cite{booch2005uml} of the tensor-conversion logic: channels depend on diagrams for the definition of their Kraus maps, during evaluation these Kraus maps are used to construct a tensor network by doubling the Kraus maps and converting them to tensors.}
  \label{fig:class-diagram}
\end{figure}

The package builds on \texttt{DisCoPy} \cite{de_felice_discopy:_2021}. Channel diagrams are built from instances of \texttt{channel.Box}, each initialised by a Kraus map within a \texttt{diagram.Diagram}. The interpretation of every \texttt{channel.Diagram} is obtained by constructing the completely positive map induced by the Kraus operators of each box and their conjugates, a process known as \emph{doubling}~\cite{coecke_categories_2016}. This procedure yields an instance of \texttt{diagram.Diagram} from each \texttt{channel.Diagram}.

Every generator in \texttt{diagram.Diagram} is itself an instance of \texttt{diagram.Box} (for example, the ZX or ZW generators), equipped with a method \texttt{truncation()} that defines the corresponding tensor by its action on basis vectors. The resulting \texttt{diagram.Diagram} can then be converted into a \texttt{discopy.tensor.Diagram} via \texttt{to\_tensor()}, and further into a \texttt{Quimb} tensor network using \texttt{to\_quimb()}. To obtain tensors of minimal dimension, an instance of \texttt{core.diagram.Diagram} is scanned from inputs to outputs so that the appropriate dimensions can be supplied to \texttt{truncation} when constructing the tensor network. This is achieved by creating ``light-cones'' for each generator to determine the number of photons which can reach it from the inputs.

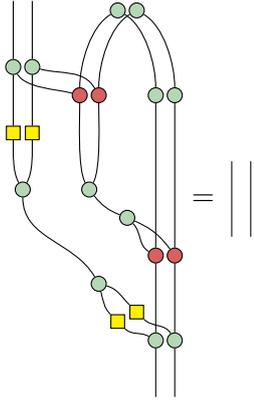
\begin{figure}[h]
  \centering
  \begin{equation*}
    \begin{tikzpicture}
	\begin{pgfonlayer}{nodelayer}
		\node [style=none] (0) at (0.5, -5.25) {};
		\node [style=none] (1) at (1, -5.25) {};
		\node [style=Z dot] (2) at (-0.5, 5) {};
		\node [style=Z dot] (3) at (0, 5) {};
		\node [style=Z dot] (4) at (0.5, -3.75) {};
		\node [style=Z dot] (5) at (1, -3.75) {};
		\node [style=X dot] (6) at (0.5, -1.5) {};
		\node [style=X dot] (7) at (1, -1.5) {};
		\node [style=Z dot] (8) at (-1, -2.25) {};
		\node [style=Z dot] (9) at (-0.25, -0.5) {};
		\node [style=Z dot] (10) at (-1.25, 0.25) {};
		\node [style=X dot] (11) at (-1.5, 2.75) {};
		\node [style=X dot] (12) at (-1, 2.75) {};
		\node [style=Z dot] (13) at (-3, 0.25) {};
		\node [style=Z dot] (14) at (-3.25, 3.5) {};
		\node [style=Z dot] (15) at (-2.75, 3.5) {};
		\node [style=none] (16) at (-3.25, 5.25) {};
		\node [style=none] (17) at (-2.75, 5.25) {};
		\node [style=hbox] (18) at (-3.25, 1.75) {};
		\node [style=hbox] (19) at (-2.75, 1.75) {};
		\node [style=hbox] (20) at (-0.5, -3.25) {};
		\node [style=hbox] (21) at (0, -3) {};
		\node [style=Z dot] (22) at (0.5, 2.75) {};
		\node [style=Z dot] (23) at (1, 2.75) {};
		\node [style=none] (24) at (1.75, 0) {$=$};
		\node [style=none] (25) at (2.5, 1) {};
		\node [style=none] (26) at (3, 1) {};
		\node [style=none] (27) at (2.5, -1) {};
		\node [style=none] (28) at (3, -1) {};
	\end{pgfonlayer}
	\begin{pgfonlayer}{edgelayer}
		\draw (0.center) to (4);
		\draw (1.center) to (5);
		\draw [in=90, out=-150, looseness=0.75] (2) to (11);
		\draw [in=90, out=-150, looseness=0.75] (3) to (12);
		\draw (4) to (6);
		\draw (5) to (7);
		\draw [in=-45, out=150] (6) to (9);
		\draw [bend right=15] (7) to (9);
		\draw [in=-90, out=120] (8) to (13);
		\draw [bend left=15] (9) to (10);
		\draw [in=-90, out=120, looseness=0.50] (10) to (11);
		\draw [in=-90, out=60, looseness=0.50] (10) to (12);
		\draw [in=-60, out=165, looseness=0.75] (11) to (14);
		\draw [in=345, out=120] (12) to (15);
		\draw (14) to (16.center);
		\draw (15) to (17.center);
		\draw (18) to (14);
		\draw [in=120, out=-90, looseness=0.75] (18) to (13);
		\draw [in=-90, out=60, looseness=0.75] (13) to (19);
		\draw (19) to (15);
		\draw [in=135, out=-75] (8) to (20);
		\draw [in=150, out=-45] (20) to (4);
		\draw [in=135, out=-15] (8) to (21);
		\draw [in=120, out=-45] (21) to (5);
		\draw [in=90, out=-30, looseness=0.75] (3) to (23);
		\draw [in=90, out=-30, looseness=0.75] (2) to (22);
		\draw (22) to (6);
		\draw (7) to (23);
		\draw (25.center) to (27.center);
		\draw (28.center) to (26.center);
	\end{pgfonlayer}
\end{tikzpicture}
  \end{equation*}
  \caption{If we double the teleportation protocol diagram from Figure~\ref{fig:teleportation-protocol-zx}, we obtain a ZX diagram which can be converted into a tensor network for evaluation or, for example, to \texttt{PyZX} for simplification. In this example, the diagram can be simplified and is equal to the identity channel on a qubit.}
  \label{fig:teleportation-protocol-zx-doubled}
\end{figure}

The backend workflow is thus:
\begin{enumerate}
  \item The user builds a channel diagram using generators from \texttt{photonic}, \texttt{qubits}, and \texttt{classical}.
  \item Calling \texttt{double()} on the channel diagram produces a doubled \texttt{core.diagram.Diagram} assembled from the Kraus maps. This is the \emph{CQ interpretation} (\emph{doubling}) of the original diagram (see Figure~\ref{fig:teleportation-protocol-zx-doubled}).
  \item The doubled diagram is composed of ZX and ZW generators.
  \item Calling \texttt{to\_tensor()} on this diagram converts it into a \texttt{DisCoPy} tensor diagram.
 \begin{itemize}
  \item Each generator is turned into a tensor by invoking its \texttt{truncation} method, which returns a tensor with the computed minimal dimensions.
 \end{itemize}

  \item The \texttt{DisCoPy} tensor diagram is translated into a \texttt{Quimb} tensor network.
  \item The resulting tensor network is ready for optimisation and execution.
\end{enumerate}

\subsection{Choosing and executing a contraction}

Once the tensors are built, their pattern of shared indices defines a tensor network.  The resulting TN is passed to \texttt{Cotengra}'s optimiser, which explores contraction trees and optionally slices indices to respect a memory cap. Slicing fixes the values of selected indices so that peak tensor sizes never exceed a bound, at the price of repeating smaller contractions; \texttt{Cotengra} balances this trade-off automatically. \texttt{Quimb} then executes the contraction according to the chosen plan, returning either a single tensor or a scalar.


\section{Examples}

We provide three examples of the use of \texttt{Optyx}: simulation of boson sampling with entangled states, distributed entanglement generation, and a variational optimisation scenario.

\subsection{Boson sampling with entangled states}
This experiment showcases exact classical simulations of quantum photonic circuits using two methods: a permanent-based approach (using a \texttt{Perceval} backend) and a tensor network-based approach (\texttt{Cotengra}, \texttt{Quimb}). Simulations were limited to 350~GB memory and 300~seconds runtime.

Circuits were constructed as layered interferometers with parameterised Mach-Zehnder (MZI) ans\"atze. For $n$ modes, three depth scalings were used: constant ($l=2$), logarithmic ($l = \lfloor \log_2(k n) \rfloor$, $k=7/5$), and linear ($l = \lfloor n/2 + 1 \rfloor$), with layer parameters sampled uniformly from $[0,1]$. Identical circuits were used for all backends and saved for reproducibility. The entangled states chosen were linear cluster states (dual-rail encoded graph states) implemented in representations compatible with both \texttt{Perceval} and \texttt{Optyx}. \emph{Monomials of number operators} \cite{Facelli:2024rhg} were used as observables with a linear scaling of the total degree with the number of photons ($\lfloor 1.5n \rfloor$).

All simulations used a shared pipeline: circuits compiled to tensor networks in \texttt{Optyx}, contraction paths optimised by \texttt{Cotengra}, and executed with \texttt{Quimb}. In \texttt{Perceval}, the corresponding unitaries were evaluated using the \texttt{NaiveBackend} implementing Ryser's \cite{ryser_combinatorial_1963} or Glynn's \cite{glynn_permanent_2010} algorithms (the choice of which is determined automatically by \texttt{Perceval}).
For an example of how an observable of a monomial operator is constructed in \texttt{Optyx}, see Code~Listing~\ref{lst:monomial-observable} in Appendix~\ref{sec:monomials-number-operators}.

\begin{figure}[h]
  \includegraphics[width=0.98\linewidth]{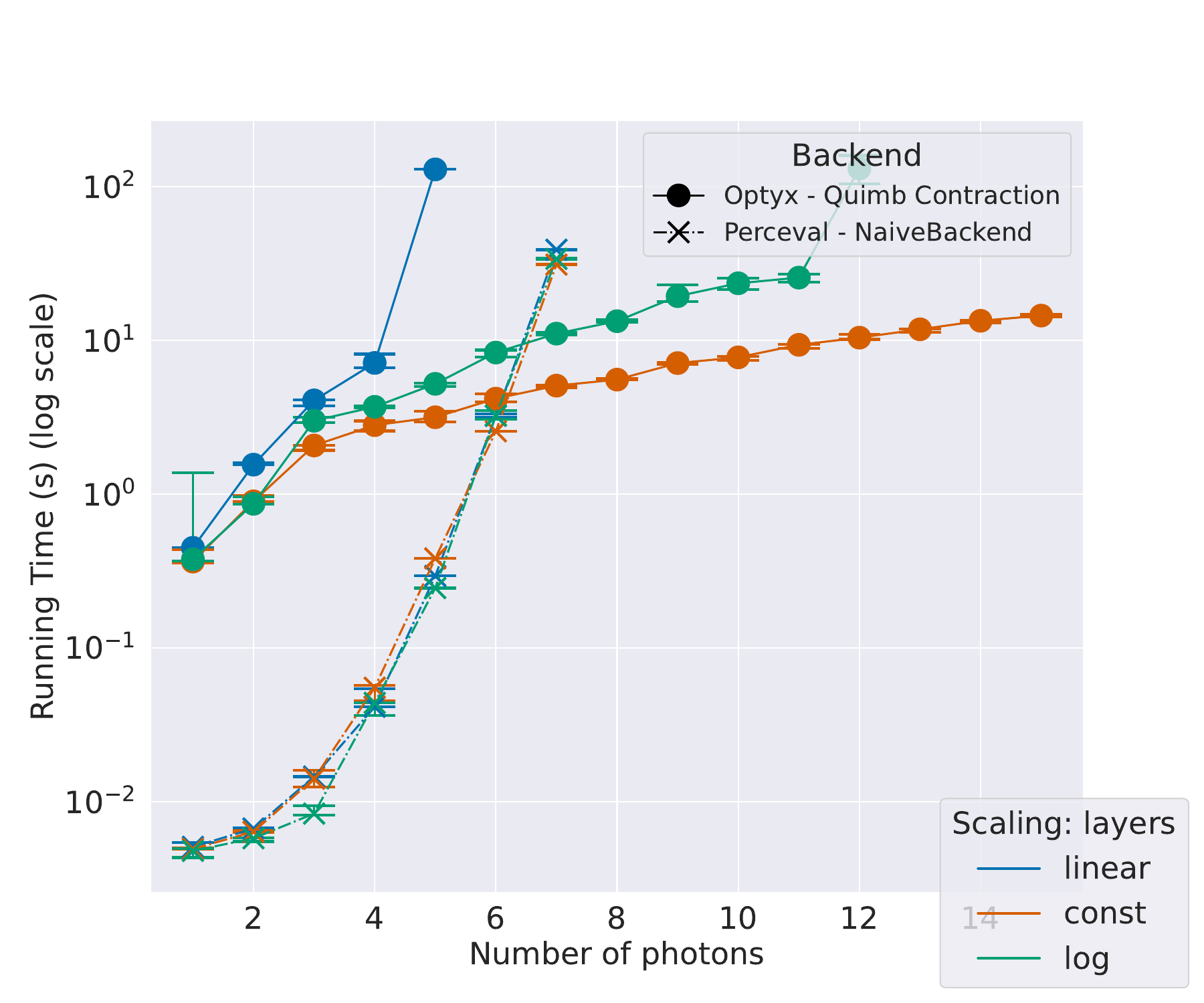}\hfill
  \caption{Exact evaluation of boson sampling observables on linear cluster states: we scale the degree of the monomial observable linearly with the number of photons ($\lfloor 1.5n \rfloor$). The depth (number of ansatz layers) is scaled linearly ($l = \lfloor n/2 + 1 \rfloor$), logarithmically ($l = \lfloor \log_2(7n/5) \rfloor$), or kept constant ($l=2$). Error bars indicate the inter-quantile range (IQR) over 5 random instances.}
  \label{fig:monomials-number-operators-lcs}
\end{figure}

For entangled inputs (Fig.~\ref{fig:monomials-number-operators-lcs}), permanent-based evaluation is fastest on very small circuits (below 6 modes), where its tractability and the overheads of tensor-network methods dominate. As circuit width and monomial degree increase, tensor-network contraction typically overtakes around 6--8 photons, especially for linearly scaling monomials where shared contractions are advantageous. At higher photon numbers, permanent methods often encounter time or memory limits, whereas tensor-network contraction extends the simulable range until large intermediate tensors exhaust memory. In this regime (moderate-high photon number, low depth, entangled sources), tensor-network methods consistently outperform Ryser/Glynn; nevertheless, both approaches become infeasible at sufficiently large scales. Permanent calculation is largely insensitive to circuit depth, while tensor-network contraction appears less sensitive to circuit width.

Overall, tensor-network methods are preferable for shallow circuits with many photons and entangled inputs. \texttt{Optyx} natively supports entangled (e.g., cluster) and hybrid qubit--photon states, which are challenging for permanent-based simulators such as \texttt{Perceval}. Preliminary results also indicate that tensor-network methods can offer advantages for observables with feed-forward dependencies, which are difficult to accommodate in permanent-based approaches. Exact tensor network contraction with \texttt{Quimb}/\texttt{Cotengra} performs better than permanent-based methods as we increase the photon number and the number of feedforward steps for low-depth circuits (sparse linear optical unitaries). While we focused on exact simulation methods, we expect advantages of tensor network over permanent based evaluation to become more prominent in approximate simulation, see e.g, Ref.~\cite{oh_classical_2021}.

\subsection{Distributed entanglement generation}
\label{sec:examples-distributed}
We simulate the effect of partial distinguishability on a distributed setup where entanglement between two remote qubits is mediated by a photonic link. This protocol has been realised experimentally, see e.g. Ref.~\cite{main2025distributed}. In each of the remote nodes, a Bell pair is created between an internal qubit and a dual-rail encoded single-photon. The photons are sent to a central ``fusion'' station performing a heralded Bell measurement: a specific click pattern projects the remote qubits into an entangled state. Being emitted by different sources, the travelling photons have internal degrees of freedom that make them partially distinguishable and this affects the fidelity of the mediated Bell pair.

To model this setup, we start with two Bell pairs on qubit registers $(A_0,A_1)$ and $(A_2,A_3)$,
\[
\ket{\Phi^+}=\tfrac{1}{\sqrt{2}}(\ket{00}+\ket{11}),
\]
\[
\rho_{\text{in}}=\ket{\Phi^+}\!\bra{\Phi^+}_{A_0A_1}\otimes \ket{\Phi^+}\!\bra{\Phi^+}_{A_2A_3}.
\]
The middle qubits $A_1, A_2$ are dual-rail encoded into photonic modes, each with a two-dimensional internal state. Let $\ket{s_1},\ket{s_2}$ be unit vectors controlling indistinguishability at fusion.

We bundle the dual-rail encoding, fusion, post-selection onto fusion success and the discarding of the four middle photonic modes (dual-rail-encoded $(A_1,A_2)$) into $ \mathcal{F}_{s_1,s_2} (\rho_{\text{in}})$ which is a heralded fusion map that takes a four-qubit input (on $A_0A_1A_2A_3$) and returns an unnormalized two-qubit state on $A_0A_3$:
\[
\tilde{\rho}_{\text{out}}(\ket{s_1},\ket{s_2})=\mathcal{F}_{s_1,s_2}\big(\rho_{\text{in}}\big).
\]
From this we report
\[
p_{\text{succ}}(\ket{s_1},\ket{s_2})=\operatorname{Tr}\!\big[\tilde{\rho}_{\text{out}}(\ket{s_1},\ket{s_2})\big],
\]
\[
\rho_{\text{out}}(\ket{s_1},\ket{s_2})=\frac{\tilde{\rho}_{\text{out}}(\ket{s_1},\ket{s_2})}{p_{\text{succ}}(\ket{s_1},\ket{s_2})},
\]
and the fidelity with the Bell target on the surviving pair,
\[
F(\ket{s_1},\ket{s_2})=\bra{\Phi^+}\,\rho_{\text{out}}(\ket{s_1},\ket{s_2})\,\ket{\Phi^+}.
\]
This (given that our target Bell state is a pure state) is equal to the channel fidelity,
\begin{multline}
\mathcal{F}(\ket{\Phi^+}\bra{\Phi^+}, \rho_{\text{out}}) = \\
\Bigl(\operatorname{Tr}\sqrt{\sqrt{\rho_{\text{out}}}\ket{\Phi^+}\bra{\Phi^+}\sqrt{\rho_{\text{out}}}}\Bigr)^2 .
\end{multline}

In the sweep we fix $\ket{s_1}$ and vary $\ket{s_2}$ so that their overlap $x=\braket{s_1 | s_2}$ ranges from $0$ (orthogonal) to $1$ (indistinguishable). We plot $F$ versus $x$: as $x\to 1$, $F\to 1$; lowering $x$ smoothly degrades the entanglement after fusion, isolating the role of internal-state overlap (see Figure~\ref{fig:distributed-entanglement}). We give the details of the construction in Appendix~\ref{sec:distributed-entanglement-generation}.

\begin{figure}[h]
  \includegraphics[width=0.98\linewidth]{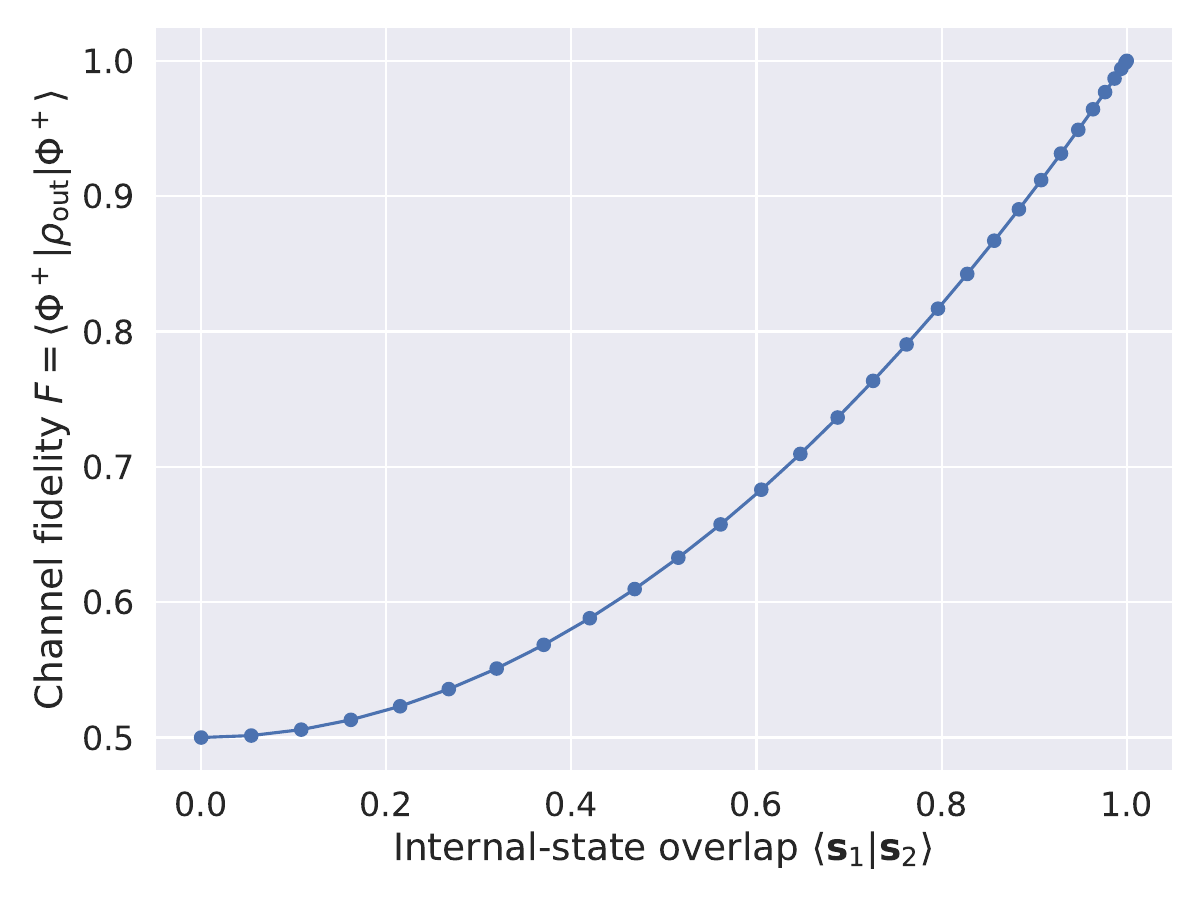}\hfill
  \caption{Distributed entanglement generation via photonic fusion with an internal degree of freedom: two Bell pairs are prepared; the middle qubits are dual-rail encoded with internal states $\ket{s_1}$ and $\ket{s_2}$, fused, and we post-select on the standard success outcome. Plotted is the post-selected fidelity $F=\bra{\Phi^+}\rho_{\text{out}}\ket{\Phi^+}$ (equal to the channel fidelity, given the target state is pure) of the surviving pair $(A_0,A_3)$ versus the internal-state overlap $x=\braket{s_1 | s_2}$. Fidelity approaches $1$ for indistinguishable internal states ($x\to 1$) and decreases smoothly as $x$ is reduced.}
  \label{fig:distributed-entanglement}
\end{figure}

\subsection{Variational eigensolver}
\label{sec:examples-variational}
The Bose-Hubbard model is a lattice model for interacting bosons, capturing the competition between tunnelling ($t$) and on-site repulsion ($U$) that drives the superfluid-Mott-insulator transition. It provides a quantitatively accurate description of cold atoms in optical lattices  \cite{fisher_boson_1989}.

For a graph $G=(V,E)$ with $|V|=N$ sites, the Hamiltonian is
\begin{equation}
\begin{aligned}
H(t,U,\mu)
&= -t \sum_{\langle i,j\rangle\in E} \big(a_i^\dagger a_j + a_j^\dagger a_i\big) \\
&\quad + \frac{U}{2}\sum_{i\in V} n_i(n_i-1) - \mu \sum_{i\in V} n_i \,,
\end{aligned}
\end{equation}
with $a_i^\dagger, a_i$ the creation/annihilation operators and $n_i=a_i^\dagger a_i$. In \texttt{Optyx}, we implement these as photonic channels and assemble $H$ by summing site- and edge-local terms; the resulting channel diagram compiles to a ZX/ZW-based intermediate representation and then to a tensor network.

\begin{figure}[h]
  \includegraphics[width=0.98\linewidth]{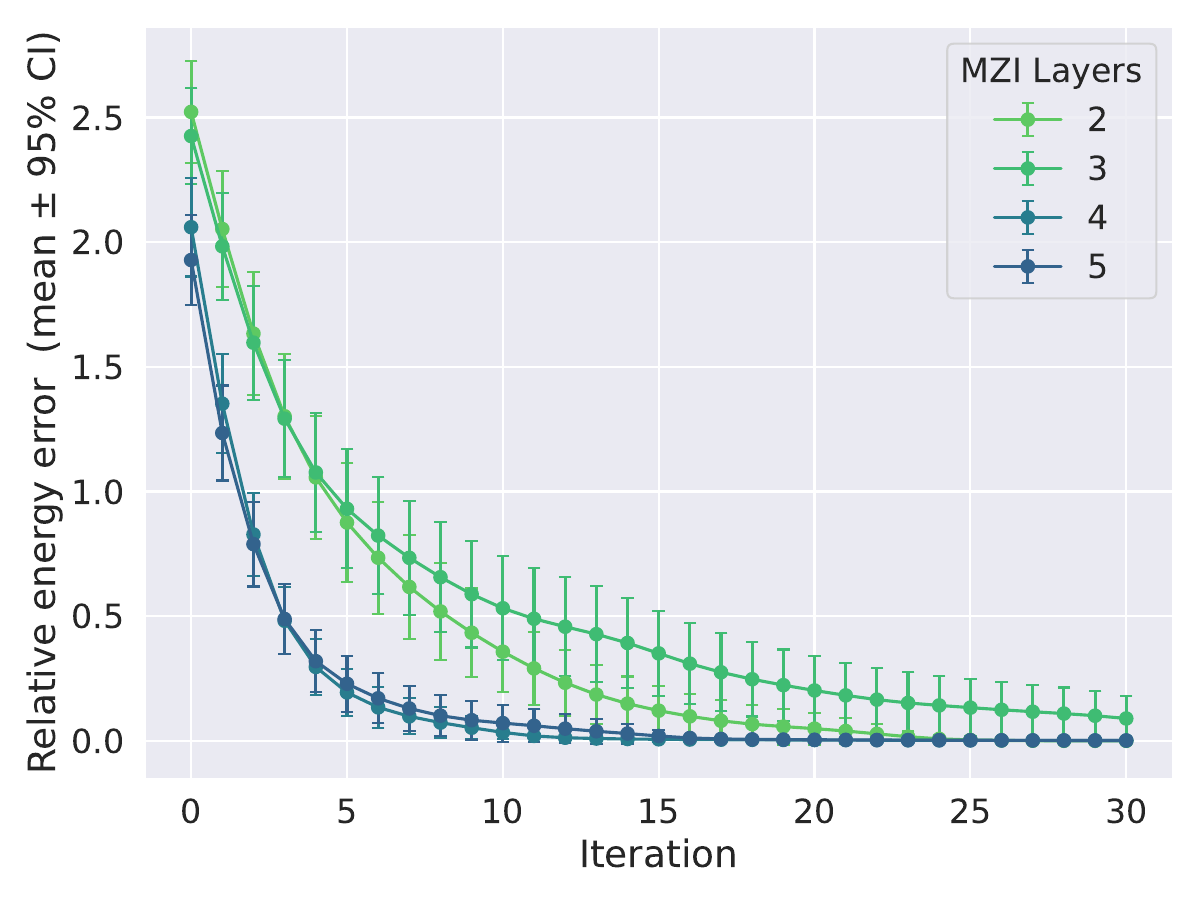}\hfill
  \caption{Bose-Hubbard model -- variational optimisation: we take a two-site chain with $(t,U,\mu)=(0.10,\,4.0,\,2.0)$, form the expectation diagram $ \bra{\psi(\boldsymbol{\theta})} H \ket{\psi(\boldsymbol{\theta})}$ for the ansatz $\ket{\psi(\boldsymbol{\theta})}$, and run gradient descent from a constant initialisation of $\boldsymbol{\theta}$. A learning-rate of $0.001$ over 30 steps produces a decrease of the energy in this small instance. 95\% Confidence interval (CI) is calculated over 100 different random initialisations. The ansatz is constructed by inputting three photons in three modes and composing them with the varying-depth MZI circuit. We also discard one mode after the ansatz to allow for variable number of photons to enter the Bose-Hubbard Hamiltonian.}
  \label{fig:bose-hubbard}
\end{figure}

We prepare a small photonic ansatz $\psi(\boldsymbol{\theta})$ and evaluate the variational energy $ E(\boldsymbol{\theta}) = \langle \psi(\boldsymbol{\theta}) | H | \psi(\boldsymbol{\theta}) \rangle$. The ansatz is based on having three photons in three modes and putting them through a variable-depth MZI circuit. We discard one mode after the MZI ansatz to allow for variable number of photons to enter the Bose-Hubbard Hamiltonian. Energies and gradients are obtained by invoking the backend to produce the corresponding results; in this example we use an exact permanent-based evaluator suitable for small instances (\texttt{PermanentBackend}), but the same diagram could compile to more scalable contraction backends. See Figure~\ref{fig:bose-hubbard} for an example of gradient descent on a two-site chain. For the construction of the Hamiltonian and the optimisation procedure, see Appendix~\ref{sec:bose-hubbard}.

\section{Conclusion and outlook}
We introduced \texttt{Optyx}, an open-source Python package for the design, simulation, and optimisation of hybrid qubit-photon quantum circuits. The framework provides a categorical semantics in which qubits, bosonic modes, and classical control are represented within a common diagrammatic language, and circuits are evaluated using tensor-network methods. This allows scalable simulation and differentiation of heterogeneous architectures, positioning \texttt{Optyx} as a platform for testing new ideas in hybrid and distributed quantum computing. The package has already been used to simulate a number of interesting scenarios. In Ref.~\cite{yin_experimental_2025}, the authors numerically simulated a photonic kernel method (a previous version of \texttt{Optyx} was used) and benchmarked SVM classification accuracies on synthetic datasets against classical kernels (Gaussian, NTK, polynomial, linear).

In the longer term, we envisage \texttt{Optyx} as a tool to probe several directions of current research. In quantum communication, the framework can be extended to model photonic carriers and their interaction with local processors, enabling the study of entanglement distribution, transmission losses, and interface fidelity. For distributed error correction, explicit qubit-photon semantics allow fault-tolerant schemes to be simulated, and the package can be adapted to facilitate the analysis of resource requirements and threshold behaviour. A tool supporting photonic nodes is needed because in quantum communication protocols, photonic errors cannot always be translated into qubit errors. Compilation for distributed hardware requires mapping abstract circuit descriptions to architectures with limited connectivity and heterogeneous links; \texttt{Optyx} can serve as a testbed for developing and benchmarking such strategies. Hybrid architectures also depend critically on light-matter interfaces. By allowing the explicit modelling of conversion processes between photons and stationary excitations, \texttt{Optyx} provides a platform for investigating the feasibility and limitations of heterogeneous systems such as ion-photonic hybrids.

On the near-term development side, several extensions are planned. Large-scale hybrid simulations will benefit from approximate tensor-network methods tailored to qubit-boson interactions, such as controlled truncation in \texttt{Quimb}, and \texttt{Optyx} offers an environment in which to design and benchmark these techniques. Improvements to the user interface and interoperability with external software stacks are also needed, so that circuits can be constructed, simulated, and exported more easily in both theoretical and experimental workflows.

In summary, \texttt{Optyx} is not only a categorical and computational framework but also a practical testbed. By extending the package in response to the demands of quantum communication, distributed error correction, compilation, and light-matter interfacing, we aim to make \texttt{Optyx} a platform where new ideas can be rapidly prototyped and evaluated against realistic architectural constraints.

\paragraph{Acknowledgements}
We would like to thank Mark Koch for developing early prototypes which laid the groundwork for \texttt{Optyx} and for providing useful feedback on the manuscript. We would also like to thank Michael Lubasch for comments and suggestions on the manuscript.

\bibliographystyle{quantum}
\bibliography{main}

\onecolumn
\appendix
\section{Kraus maps and custom channels}
\label{sec:kraus_maps}
\texttt{Optyx} provides a way to define custom channels and boxes, which can be used to create new generators. In order to make the new generator simulable using tensor networks, a box inheriting from \texttt{core.diagram.Box} must implement the \texttt{truncation\_specification} method (for the class diagram see Figure~\ref{fig:class-diagram}). This specifies how a box acts on the basis vectors of the input and output edges, and is used to construct a tensor for a given box. Once the box representing a tensor is built, it can be used as a Kraus map for a channel. More details of this are explained in Section~\ref{sec:construction-of-tensor-networks} and in the documentation of \texttt{Optyx}.

\begin{lstlisting}[language=Python,
  caption={Template for defining a custom box and channel in \texttt{Optyx}.},
  label={lst:custom-channel}]
from optyx.core.diagram import Box
from optyx.core.channel import Channel
from optyx.utils.utils import BasisTransition

class MyCustomBox(Box):
  def __init__(self, dom, cod, myparam):
    super().__init__("MyBox", dom, cod)
    self.myparam = myparam

  def truncation_specification(
      self,
      inp: Tuple[int, ...] = None,
      max_output_dims: Tuple[int, ...] = None
  ) -> Iterable[BasisTransition]:
    # get output basis_elements
    # for an input basis element inp
    for basis_element in basis_elements:
      # calculate the amplitude
      yield BasisTransition(
          out=basis_element,
          amp=amplitude
      )

mychannel = lambda dom,cod,mp: Channel(
  "MyChannel",
  MyCustomBox(dom, cod, mp)
)
\end{lstlisting}

\section{Teleportation using fusion measurements}
\label{sec:teleportation-fusion}
\begin{figure}[h]
  \centering
  \includegraphics[width=0.5\linewidth]{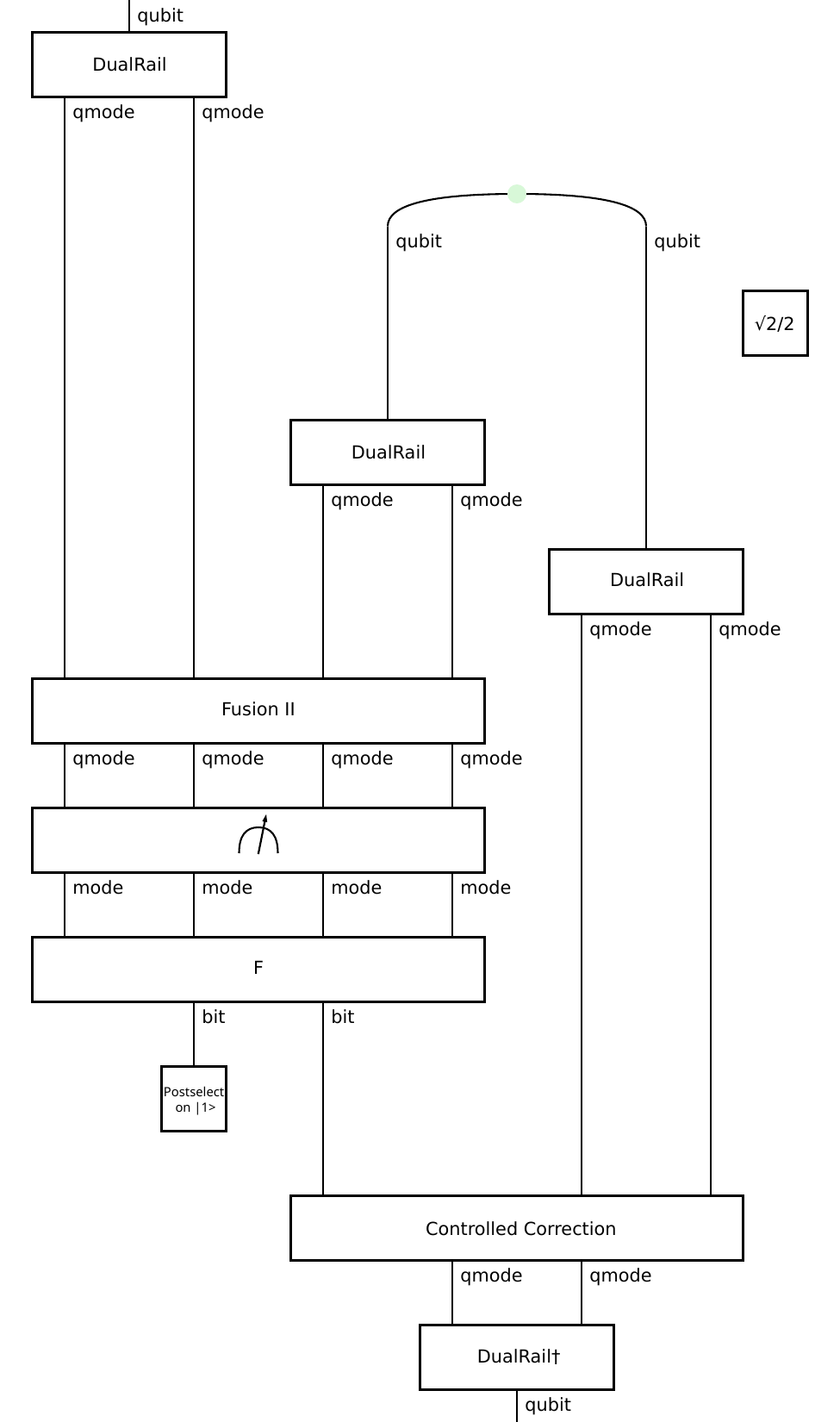}
  \caption{Diagram of the teleportation protocol using a fusion measurement. The protocol demonstrates the transfer of quantum information using entanglement and classical communication.}
  \label{fig:teleportation-protocol}
\end{figure}

 The example in Code~Listing~\ref{lst:teleportation-fusion} shows how to implement a teleportation protocol using a fusion type II measurement. The protocol uses a Bell pair as a channel to teleport a qubit from one dual-rail encoding to another. The measurement result is used to apply a correction on the teleported qubit. The entire protocol is shown in Figure~\ref{fig:teleportation-protocol}. We can verify that the protocol implements the identity channel (up to a normalisation factor) by evaluating the diagram and comparing it to the identity channel on one dual-rail qubit.

This protocol showcases the seamless integration of classical and quantum resources in a hybrid photon-qubit architecture in \texttt{Optyx}.

\begin{lstlisting}[language=Python,
  caption={Teleportation protocol using dual-rail encoding, fusion measurement, and classical control.},
  label={lst:teleportation-fusion}]
correction = BitControlledGate(
    HadamardBS() >>
    (Phase(0.5) @ qmode) >>
    HadamardBS()
)

channel_bell = (
    Z(0, 2) @ Scalar(0.5**0.5) >> DualRail(1) @ DualRail(1)
)

teleportation = (
  DualRail(1) @ channel_bell >>
  FusionTypeII() @ qmode**2 >>
  PostselectBit(1) @ correction >>
  DualRail(1).dagger()
)

array_teleportation = teleportation.eval().tensor.array

array_id = (
    Id(1) @ Scalar(0.5**0.5)
).eval().tensor.array

assert np.allclose(array_teleportation, array_id)
\end{lstlisting}

\section{Monomials of number operators}
\label{sec:monomials-number-operators}
We can create a circuit to evaluate a monomial of number operators in a photonic circuit. This example uses a bosonic product state as input (\texttt{Create(1, 1, 1, 1)}) but other states can be considered.

\begin{lstlisting}[language=Python,
  caption={
Builds and evaluates a monomial of number operators in a photonic circuit. A random unitary $U$ is generated by a rectangular MZI mesh \texttt{ansatz(w,l)} with uniformly sampled parameters. For \texttt{powers} $=\{p_k\}$, the layer implements $\bigotimes_k n_k^{p_k}$ by applying \texttt{NumOp()} $p_k$ times on mode $k$. The diagram computes
$\langle \psi \vert\, U^\dagger \!\left(\bigotimes_k n_k^{p_k}\right)\! U \,\vert \psi \rangle$.
  },
  label={lst:monomial-observable}]
from optyx.photonic import (
    ansatz, NumOp, Create
)

def chip_mzi(w, l):
    """Create a random unitary
    using a rectangular mesh of MZIs."""
    ansatz_ = ansatz(w, l)
    symbs = list(ansatz_.free_symbols)
    s = [(i, np.random.uniform(0, 1))
         for i in symbs]
    return ansatz_.subs(*s)

powers = [1, 2, 1, 3]
U = chip_mzi(4, 4)
monomial_layer = Diagram.tensor(
    *[
        Diagram.then(
            *[NumOp()]*p
        ) for p in powers
    ]
)

diagram = (
    Create(1, 1, 1, 1) >>
    U >>
    monomial_layer >>
    U.dagger() >>
    Create(1, 1, 1, 1).dagger()
)

result = diagram.eval().tensor

\end{lstlisting}

\section{Distributed entanglement generation}
\label{sec:distributed-entanglement-generation}
We simulate fusion-based entanglement generation with an internal degree of freedom (DoF) that controls distinguishability. Two Bell pairs are prepared; the two middle qubits are mapped to dual-rail photonic modes with internal states $\ket{s_1},\ket{s_2}$. These modes interfere in a fusion primitive and we herald success by post-selecting detector outcomes $(1,0)$. Upon success, the photonic modes are discarded and the two outer qubits remain.

In the sweep, $\ket{s_1}$ is fixed and $\ket{s_2}$ is rotated from parallel to orthogonal, so the overlap $\braket{s_1 | s_2}$ ranges from $1$ to $0$. For each setting we compute:

\begin{itemize}
\item \textbf{Numerator (overlap with Bell):} the overlap of the post-selected output with $\ket{\Phi^+}$,
\[
F(\ket{s_1},\ket{s_2}) \;=\; \big\langle \Phi^+ \big| \,\tilde{\rho}_{\text{out}}(\ket{s_1},\ket{s_2})\, \big| \Phi^+ \big\rangle.
\]
\item \textbf{Denominator (success probability):} the post-selected trace,
\[
p_{\text{succ}} \;=\; \operatorname{Tr}\!\left[\,\tilde{\rho}_{\text{out}}(\ket{s_1},\ket{s_2})\,\right].
\]
\item \textbf{Fidelity (conditional on success):}
\[
F \;=\; \frac{F(\ket{s_1},\ket{s_2})}{p_{\text{succ}}} \;=\; \bra{\Phi^+}\,\rho_{\text{out}}(\ket{s_1},\ket{s_2})\,\ket{\Phi^+},
\qquad
\rho_{\text{out}} \;=\; \frac{\tilde{\rho}_{\text{out}}}{p_{\text{succ}}}.
\]
\end{itemize}

\begin{lstlisting}[language=Python,
  caption={Distributed entanglement generation via photonic fusion with an internal degree of freedom.},
  label={lst:distributed-entanglement}]
from optyx.qubits import Z, Scalar, Id, Discard
from optyx.photonic import DualRail, FusionTypeII
from optyx.classical import PostselectBit
from optyx.core.channel import Diagram
import numpy as np, math
import matplotlib.pyplot as plt

bell_state = (Z(0, 2) @ Scalar(0.5 ** 0.5))
dual_rail_encoding = lambda s: DualRail(1, internal_states=[s])
post_select = PostselectBit(1) @ PostselectBit(0)

def fusion(internal_state_1, internal_state_2):
    @Diagram.from_callable(dom=0, cod=2)
    def d():
        a = (bell_state @ bell_state)()
        b = (dual_rail_encoding(internal_state_1) @ dual_rail_encoding(internal_state_2))(a[1], a[2])
        c = FusionTypeII()(*b)
        post_select(c[0], c[1])
        return a[0], a[3]
    return d

def rotated_unit_vectors(n=30):
    for i in range(n):
        theta = i * (math.pi/2) / (n - 1)
        yield (math.cos(theta), math.sin(theta))

internal_state_1 = (1.0, 0.0)
unit_vectors = list(rotated_unit_vectors(30))

xs, Fs = [], []
for s2 in unit_vectors:
    encoding = dual_rail_encoding(internal_state_1) @ dual_rail_encoding(s2)
    experiment = (bell_state @ bell_state
                  >> Id(1) @ (encoding >> FusionTypeII() >> post_select) @ Id(1))
    num = (experiment >> bell_state.dagger()).inflate(2).eval().tensor.array
    den = (experiment >> Discard(2)).inflate(2).eval().tensor.array
    xs.append(np.inner(s2, internal_state_1))
    Fs.append(num / den)
\end{lstlisting}

\section{Bose-Hubbard model}
\label{sec:bose-hubbard}

We construct the Bose--Hubbard Hamiltonian as a sum of terms over a register of
$N=|V(G)|$ bosonic modes. Given a \texttt{NetworkX} graph $G=(V,E)$ and scalars $t,\mu,U$, the routine
\texttt{bose\_hubbard\_from\_graph} returns a single \texttt{Diagram} $H$ built entirely in
function syntax. For every undirected edge $\{i,j\}\in E$ we add the hopping pair
$-t(a_i^\dagger a_j + a_j^\dagger a_i)$ by defining
\texttt{creation\_op} on wire $i$ and \texttt{annihilation\_op} on wire $j$. For each site
$i\in V$ we add the on-site interaction $(U/2)\,a_i^\dagger a_i^\dagger a_i a_i$
(i.e. $(U/2)\,n_i(n_i-1)$ with $n_i=a_i^\dagger a_i$) by composing two creations followed by two
annihilations on wire $i$ and multiplying by \texttt{Scalar(U/2)}. Finally, we add the chemical
potential term $-\mu\,n_i$ using the provided one-mode number operator \texttt{NumOp()} with
\texttt{Scalar(-mu)}. All contributions are summed into $H$. See Code~Listing~\ref{lst:bose-hubbard}.

\begin{lstlisting}[language=Python,
  caption={Bose-Hubbard model Hamiltonian for a graph.},
  label={lst:bose-hubbard}]
import networkx as nx
from optyx import Diagram, qmode
from optyx.photonic import NumOp, Scalar
from optyx import Channel
from optyx.core.diagram import mode
from optyx.core.zw import W, Create, Select

creation_op = Channel(
    "a^dagger",
    Create(1) @ mode >> W(2).dagger()
)

annihilation_op = Channel(
    "a",
    W(2) >> Select(1) @ mode
)

def bose_hubbard_from_graph(graph: nx.Graph, t: float, mu: float, U: float):
    nodes = sorted(graph.nodes())
    idx = {u: i for i, u in enumerate(nodes)}
    N = len(nodes)

    H = None

    # Hopping: -t (a_i^dagger a_j + a_j^dagger a_i)
    for u, v in graph.edges():
        i, j = idx[u], idx[v]
        for src, dst in ((i, j), (j, i)):
            @Diagram.from_callable(dom=qmode**N, cod=qmode**N)
            def hop(*in_wires, src=src, dst=dst):
                out = list(in_wires)
                out[src] = creation_op(out[src])
                out[dst] = annihilation_op(out[dst])
                Scalar(-t)()
                return tuple(out)
            H = hop if H is None else (H + hop)

    # On-site: (U/2) a_i^dagger a_i^dagger a_i a_i
    for u in nodes:
        i = idx[u]
        @Diagram.from_callable(dom=qmode**N, cod=qmode**N)
        def quartic_i(*in_wires, i=i):
            w = creation_op(creation_op(in_wires[i]))
            w = annihilation_op(annihilation_op(w))
            out = list(in_wires); out[i] = w
            Scalar(U/2)()
            return tuple(out)
        H = quartic_i if H is None else (H + quartic_i)

    # Chemical: -μ n_i
    for u in nodes:
        i = idx[u]
        @Diagram.from_callable(dom=qmode**N, cod=qmode**N)
        def n_i(*in_wires, i=i):
            out = list(in_wires)
            out[i] = NumOp()(out[i])
            Scalar(-mu)()
            return tuple(out)
        H = n_i if H is None else (H + n_i)

    return H
\end{lstlisting}

We use a photonic variational ansatz circuit consisting of three layers of Mach-Zehnder interferometers and post-selection on one mode, as shown in Code~Listing~\ref{lst:bose-hubbard-ansatz}.

\begin{lstlisting}[language=Python,
  caption={Bose-Hubbard variational ansatz circuit.},
  label={lst:bose-hubbard-ansatz}]
from optyx import photonic, classical

circuit = photonic.Create(1, 1, 1) >> photonic.ansatz(3, 4)
\end{lstlisting}

We instantiate a two-site chain $G=\texttt{path\_graph}(2)$ with parameters $(t,U,\mu)=(0.10,\,4.0,\,2.0)$ and construct the Bose--Hubbard Hamiltonian diagram $H$. Given a defined variational expectation diagram $\texttt{expectation}=\langle\psi(\boldsymbol{\theta})|\,H\,|\psi(\boldsymbol{\theta})\rangle$, the code uses the parameters $\boldsymbol{\theta}$ and defines a callable energy $E(\boldsymbol{\theta})$.
Gradients are obtained by applying \texttt{expectation.grad(s)} for each parameter $s \in \boldsymbol{\theta}$ to yield $\nabla E(\boldsymbol{\theta})$.

A simple gradient-descent routine \texttt{optimize} initializes $\boldsymbol{\theta}^{(0)}=[2,\dots,2]$, uses an initial learning rate $\ell_{0}=5$, and runs $12$ iterations. Each step updates coordinates,
\[
\theta_k \leftarrow \theta_k - \ell\,(\nabla E(\boldsymbol{\theta}))_k \quad (k=1,\dots,m).
\]
 See Code~Listing~\ref{lst:bose-hubbard-optimization}.

\begin{lstlisting}[language=Python,
  caption={Variational optimisation of a two-site Bose-Hubbard model.},
  label={lst:bose-hubbard-optimization}]
import networkx as nx
from tqdm import tqdm
from optyx.core.backends import PermanentBackend

graph = nx.path_graph(2) # 2 sites
t, U, mu = 0.10, 4.0, 2.0

# one extra qmode to allow variable numbers
# of photons to enter the circuit of the bose-hubbard Hamiltonian
hamiltonian = bose_hubbard_from_graph(graph, t, mu, U) @ qmode

def to_float(x):
    if isinstance(x, complex):
        assert x.imag < 1e-8, x
        return x.real
    return x

free_syms = list(expectation.free_symbols)

f_exp = lambda xs: to_float(
    expectation.lambdify(*free_syms)(*xs)
    .eval(PermanentBackend())
    .tensor
    .array
)

def d_f_exp(xs):
    return [
        expectation.grad(s).lambdify(*free_syms)(*xs)
        .eval(PermanentBackend())
        .tensor
        .array
        for s in free_syms
    ]

def optimize(x0):
    x = x0
    lr = 0.001
    steps = 30

    xs = []
    fxs = []
    dfxs = []

    for _ in tqdm(range(steps)):
        fx = f_exp(x)
        dfx = d_f_exp(x)

        xs.append(x[::])
        fxs.append(fx)
        dfxs.append(dfx)
        for i, dfxx in enumerate(dfx):
            x[i] = to_float(x[i] - lr * dfxx)


    xs.append(x[::])
    fxs.append(f_exp(x))
    dfxs.append(d_f_exp(x))

    return xs, fxs, dfxs

xs, fxs, dfxs = optimize([2]*len(free_syms))

\end{lstlisting}

\end{document}